\RequirePackage{fix-cm}

\documentclass[twocolumn,epjc3]{svjour3}

\smartqed  
\RequirePackage{graphicx}
\usepackage{amsmath}
\usepackage{slashed}
\usepackage{subcaption}
\usepackage{amssymb}
\usepackage{bm}
\usepackage{cuted}
\usepackage{lineno}
\usepackage{hyperref}
\journalname{Eur. Phys. J. A}
\begin{document}

\title{Mapping the Electromagnetic Fields of Heavy-Ion Collisions with the Breit-Wheeler Process 
}
\titlerunning{Mapping the Electromagnetic Fields of Heavy-Ion Collisions}        

\author{J. D. Brandenburg\thanksref{e1,addr1,addr2}
        \and
        W. Zha\thanksref{e2,addr3} 
        \and
        Z. Xu \thanksref{e3,addr1}
}

\thankstext{e1}{e-mail: jbrandenburg@bnl.gov}
\thankstext{e2}{e-mail: first@ustc.edu.cn}
\thankstext{e3}{e-mail: xzb@bnl.gov}


\institute{Brookhaven National Laboratory, Upton NY 11973-5000 \label{addr1}
           \and
           Center for Frontiers in Nuclear Science affiliate at Stony Brook University, Stony Brook NY, 11794-38000 \label{addr2}
           \and
           University of Science and Technology of China, Hefei, China\label{addr3}
}

\date{Received: 30-March-2021 / Accepted: 29-September-2021}

\maketitle

\begin{abstract}
Ultra-relativistic heavy-ion collisions are expected to produce the strongest electromagnetic fields in the known Universe. 
These highly-Lorentz contracted fields can manifest themselves as linearly polarized quasi-real photons that can interact via the Breit-Wheeler process to produce lepton anti-lepton pairs. The energy and momentum distribution of the produced dileptons carry information about the strength and spatial distribution of the colliding fields. Recently it has been demonstrated that photons from these fields can interact even in heavy-ion collisions with hadronic overlap, providing a purely electromagnetic probe of the produced medium. In this review we discuss the recent theoretical progress and experimental advances for mapping the ultra-strong electromagnetic fields produced in heavy-ion collisions via measurement of the Breit-Wheeler process. 
\end{abstract}

\setcounter{tocdepth}{3}
\tableofcontents
%
%
\section{Introduction}
\label{intro}
Heavy-ion collisions are expected to produce the strongest electromagnetic (EM) fields in the known Universe, of order $10^{15}$ Tesla~\cite{Norbeck2006,skokovEstimateMagneticField2009,mclerranCommentsElectromagneticField2014}.  In both Quantum Electrodynamics (QED)~\cite{battestiHighMagneticFields2018,hattoriVacuumBirefringenceStrong2013} and Quantum Chromodynamics (QCD)~\cite{asakawaElectricChargeSeparation2010,skokovChiralMagneticEffect2017}, such strong EM fields are expected to lead to various exotic phenomena and have therefore been a source of intense study over the decades. Recently, there has been a large amount of experimental and theoretical interest in emergent magnetohydrodynamical QCD phenomena~\cite{kharzeevChiralMagneticVortical2015} that may arise in heavy-ion collisions (HIC). Specifically, the Chiral Magnetic Effect (CME)~\cite{fukushimaChiralMagneticEffect2008a,kharzeevEffectsTopologicalCharge2007}, an anomalous electric current driven by chiral imbalances, has been searched for and studied extensively~\cite{buividovichNumericalEvidenceChiral2009,wangSearchChiralMagnetic2018,zhaoExperimentalSearchesChiral2019}. However, the realization of such phenomena requires the presence of extremely strong electromagnetic fields simultaneous with the production of a quark-gluon plasma (QGP). While heavy-ions are expected to produce sufficiently strong fields for these exotic phenomena, previously there has not been any clear experimental measurement of the field strengths. 

While the primary interest in high-energy heavy-ion collisions is the production of a quark gluon plasma, the high $Z$ nuclei also provide an ideal laboratory for studying photon mediated process due to the $\propto Z^2$ enhancement of coherent quasi-real photons which comprise the highly-Lorentz contracted EM fields. 
Exclusive photon mediated processes, especially the $\gamma\gamma \rightarrow l^+l^-$ processes (hereinafter $\gamma\gamma$ process), have been studied in so-called ultra-peripheral collisions (UPC) for some time~\cite{abbasCharmoniumPairPhotoproduction2013a,bertulaniElectromagneticProcessesRelativistic1988,baurElectronPositronPairProduction2007a,starcollaborationProductionEnsuremathPairs2004}. In ultra-peripheral collisions the nuclei pass one another with an impact parameter $b$ large enough that no nuclear overlap occurs, roughly for $b>2R_A$, with $R_A$ being the nuclear radius. 
It was realized only recently that such $\gamma\gamma$ processes can occur and even be observed in head-on heavy-ion collisions. The STAR~\cite{starcollaborationLowEnsuremathPair2018b}, ATLAS~\cite{atlascollaborationObservationCentralityDependentAcoplanarity2018a} and ALICE~\cite{lehnerDielectronProductionLow2019a} collaborations have demonstrated that the $\gamma\gamma$ processes can be statistically isolated in semi-central to central collisions from the ``underlying event'' produced by the hadronic interactions. 
These measurements also reported significant broadening of the dilepton transverse momentum ($P_\perp$) distributions compare to the same processes in UPCs, leading to speculation on the mechanism responsible for the modifications. 

The unsuccessful description of the STAR data by STARlight led to the attribution of the $P_\perp$ broadening to the possible residual magnetic field trapped in an electrically conducting QGP~\cite{mclerranCommentsElectromagneticField2014}. 
Similarly, the ATLAS collaboration qualified the effect via the acoplanarity of lepton pairs $\alpha$ in contrast to the measurement in UPC and explained the additional broadening through multiple electromagnetic scatterings in the hot and dense medium. 
However, each of these descriptions of the broadening effect assume that there is no impact parameter dependence of the transverse momentum distribution for the electromagnetic production of lepton pairs.
Additional measurements from STAR~\cite{starcollaborationMeasurementMomentumAngular2021} and CMS~\cite{cmscollaborationObservationForwardNeutron2020a} have since demonstrated that significant $P_\perp$ broadening results from the impact parameter dependence of the process, even in the absence of a quark gluon plasma or other medium. 
This impact parameter dependence can be understood in terms of the spatial distribution of the colliding fields and therefore provides an avenue for experimentally measuring the strength and spatial distribution (i.e. ``mapping'') of the strong EM fields for the first time.
Furthermore, STAR~\cite{starcollaborationMeasurementMomentumAngular2021} has measured angular modulations in the daughter lepton momentum with respect to the dilepton pair momentum, experimentally confirming that the colliding photons are linearly polarized.
Since the photon polarization is defined in terms of the radial electric field emanating from the nucleus, with magnetic field circling it, detailed measurement of the photon polarization effects provides a new tool for the discovery of vacuum birefringence and a method for mapping the two dimensional distribution of the transverse EM fields.

An understanding of the physics involved in the $\gamma\gamma$ processes is necessarily based on a theoretical or phenomenological framework. Over the last few decades there have existed two main frameworks: equivalent photon approximation (EPA) and lowest-order QED (LOQED). 
The equivalent photon approximation has been used for many decades to simplify the calculation of photon mediated processes.
For instance, the two-photon QED processes in high-energy $e^+e^-$ collisions~\cite{brodskyTwoPhotonMechanismParticle1971a,budnev_two-photon_1975} can be formulated in terms of the collision of Weizsäcker-Williams photons as summarized by the Particle Data Group~\cite{zyla_review_2020} (section 50.7).
Specifically, Eq.~50.44 of Ref.~\cite{zyla_review_2020} exemplifies the use of the EPA in $e^+e^-$ collisions.  
In this picture, the photons are assumed to be emitted from the centers of the incoming electron (positron) and then collide. 
It requires minimum and maximum virtuality with a finite four-momentum transfer ($-k^2$) for the photons to be emitted from the electron and positron (see Eq.~50.44 to Eq.~50.45 in Ref.~\cite{zyla_review_2020}). 
This is necessary because the photon flux in such a formalism diverges at both high and low photon $k^2$. 
Such an approach has been adapted in models for ultra-peripheral heavy-ion collisions with the exclusion of the field inside the nucleus~\cite{kleinSTARlightMonteCarlo2017b,kleinExclusiveVectorMeson1999a} (such as STARlight, which has been commonly used for comparison with high-energy experiments for the last two decades) or with the addition of nuclear charge form factors~\cite{klein_two-photon_2018}.
In all EPA calculations for ultra-peripheral heavy-ion collisions the Breit-Wheeler two-photon cross section (applicable only for real photons) has been used. 
This approach can describe the cross section reasonably well with the decay kinematics (specifically the the polar angular distribution) computed in an ad hoc way by using the angular distribution predicted by LOQED for the collision of real photons~\cite{brodskyTwoPhotonMechanismParticle1971a,kleinSTARlightMonteCarlo2017b}. 

However, there are fundamental differences in the assumptions used to implement the EPA in an $e^+e^-$ collider versus in heavy-ion UPCs. 
While the EPA formalism for an $e^+e^-$ collider explicitly requires deflection of the scattering electron (positron) beam particles leading to minimum and maximum four-momentum transfer (finite virtuality)~\cite{budnev_two-photon_1975}, the EPA implementation traditionally used to describe UPCs is formulated in terms of the external field approximation, which requires the colliding heavy ions to maintain their straight-line trajectory both before and after the collision~\cite{PhysRevC.47.2308}.
Unlike in the case for an $e^+e^-$ collider, the photon flux does not diverge in UPCs because the low-energy photon flux is regulated by the finite Lorentz factor of the ions ($k^2 \ge (\omega/\gamma)^2~ \gtrsim ~(2$ MeV$)^2$), where $\omega$ is the photon energy and $\gamma$ is the ion Lorentz factor) and the high-energy photon flux is naturally cut off by the finite size of the ion's charge distribution ($k^2~\lesssim~(1/R_A)^2 \simeq (30$ MeV$)^2$, $R_A$ is the nuclear radius)(e.g. Eq. (38)$-$(45) in Ref.~\cite{PhysRevC.47.2308}). 
Similarly, in the high-power laser-driven nonlinear Breit-Wheeler process~\cite{SLACPhysRevLett.79.1626}, the photon generated by the electron-laser collisions serves as an intermediate propagator and its divergence is cutoff by the finite duration of the laser pulse~\cite{SLACTridentQEDPRL2010} with a laser pulse length about 10 times that of the laser photon wavelength.
While the implemented EPA approaches have provided a relatively good description of the measured $\gamma\gamma$ process in UPCs over the years, they have been unable to describe various experimental results, such as the recently observed broadening of the pair transverse momentum with impact parameter. 
While impact parameter dependence in relativistic heavy-ion collisions of the cross section has been implemented with additional factors, this kind of implementation fails to describe the transverse-momentum spectra and fails to produce any impact parameter dependence on the spectra as observed in experiments. 
Additionally, the traditional EPA approach ignores the spin-momentum correlation of the photons. 
These aspects of the process kinematics are especially important, since, if the photon transverse momentum is determined by the uncertainty principle and virtuality, it would not depend on collision impact parameter nor would it correlate with the orientation of the Coulomb field. 
This deficiency has been pointed out in recent theoretical attempts to produce a generalized EPA (gEPA) framework for heavy-ion collisions~\cite{hencken_electromagnetic_1994,zhaInitialTransversemomentumBroadening2020b}. Additionally, the traditional EPA approach is also unable to provide any azimuthal dependence resulting from the photon spin.
Though not internally consistent, an approximate azimuthal dependence can be achieved in the traditional EPA approach by introducing the transverse-momentum distribution (TMD) from the Poynting vector of the field as is done in Ref.~\cite{liProbingLinearPolarization2019} and similarly in SuperChic3~\cite{SuperChic3}.

In addition to the phenomenological approaches using the traditional EPA method, the $\gamma\gamma$ processes may also be computed via Lowest-Order QED Feynman diagrams~\cite{hencken_electromagnetic_1994,Hencken:1995me,zhaInitialTransversemomentumBroadening2020b,liImpactParameterDependence2020}. 
In this case, the semi-classical electromagnetic fields resulting from the Lorentz-boosted Coulomb field are used as input to the lowest order Feynman diagram calculations. 
Some models~\cite{hencken_electromagnetic_1994,zhaInitialTransversemomentumBroadening2020b} directly use the full four-momentum vector generated by the field as input to the (real photon) vertex function while others~\cite{PhysRevC.47.2308,liImpactParameterDependence2020} use the real-photon and its three-momentum vector instead. 
Many comparisons have been carried out via analytic formula~\cite{PhysRevC.47.2308}, numerical calculations between gEPA and LOQED with virtuality (at $b=0$)~\cite{hencken_electromagnetic_1994}, and LOQED with and without virtuality~\cite{zhaInitialTransversemomentumBroadening2020b,liImpactParameterDependence2020}. All these have shown that the difference in the physical observables, including the transverse-momentum distributions, between full four-momentum vector input and real-photon input is negligible in RHIC and LHC kinematics (e.g. see the derivation of the exact form of the matrix element for $\gamma\gamma \rightarrow l^+l^-$ process in Eq.~28-33 from Ref.~\cite{hencken_electromagnetic_1994}). 
Specifically, Eq.~25 in Ref.~\cite{PhysRevC.47.2308} shows that in the transition current, the term which dominates is the first term with the magnitude of the transverse component of the Poynting vector ($k_\perp$) at about a factor of $1/\gamma$ times its longitudinal magnitude  ($k_\perp\simeq\omega/\gamma$). 
Neglecting the other terms which are suppressed by $1/\gamma^2$, one can arrive at the vertex function for real (massless) photons interactions  (i.e. Eq.~28 in Ref.~\cite{PhysRevC.47.2308}) and can consequently justify the real-photon approximation for all the physics observables in the specified kinematic range. Throughout this review, the colliding photons that satisfy the above conditions are referred to as ``quasi-real'', meaning that they behave as real photons with respect to all final state observables since their virtuality, though not exactly zero, is inconsequential even for $e^+e^-$ pair production (since electrons have the lightest mass of all leptons, any photon virtuality would have the largest impact on $e^+e^-$ production). Breit and Wheeler first introduced the concept of pair production from the fusion of two real photons~\cite{Breit-wheeler1934zz}, i.e. the $:\gamma\gamma \rightarrow e^+e^-$ process. In this review we use ``Breit-Wheeler process'' to refer to the family of $\gamma\gamma \rightarrow l^+l^-$ processes where $l=e, \mu, \tau$. 

There have been several reviews of the photon-photon process in heavy ion collisions~\cite{baurElectronPositronPairProduction2007a,baltzPhysicsUltraperipheralCollisions2008,brodskyHighEnergyPhotonphoton1995,baurCoherentPhotonphotonProcesses1993a} focusing on opportunities available at RHIC and LHC, including some that discuss recent developments, e.g. ~\cite{kleinPhotonuclearTwoPhotonInteractions2020}. 
In this review we focus on the connection between the Breit-Wheeler process and the EM fields of the ultra-relativistic heavy-ions. 
The primary focus of this review is to discuss recent developments that demonstrate the viability of using $\gamma\gamma$ interactions to experimentally map the electromagnetic fields produced by heavy-ion collisions in order to provide novel input to fundamental investigations in both QED and QCD. 
This review is structured as follows: First we introduce the concept of pair production in vacuum by strong fields followed by a brief review of the equivalent photon approximation. Next we discuss the impact parameter dependence of the dilepton kinematics, followed by a discussion of the photon polarization effects. Finally, we close the paper with discussion on the experimental mapping of the strong EM fields, and end with a brief discussion of some future prospects.

\section{The $\gamma\gamma \rightarrow l^+l^-$ Process}
\label{sec:process}
The concept of $e^+e^-$ pair production from the vacuum in the presence of a uniform and constant electric field was first introduced by Sauter~\cite{sauterUeberVerhaltenElektrons1931} and later developed by Euler and Heisenberg~\cite{heisenbergFolgerungenAusDiracschen1936}. Finally, the paradigm of $e^+e^-$ pair production in Quantum Electrodynamics was developed by Schwinger~\cite{schwingerGaugeInvarianceVacuum1951}. The Schwinger mechanism, applicable for constant and spatially uniform electric fields, leads to pair production above a critical electric field of
\begin{equation}
E_c = \frac{m^2c^3}{e\hbar}  \simeq 1.3\times 10^{16}\ {\rm V/cm}.
\end{equation}
For comparison, the order of magnitude estimation for the maximum achieved electric field in heavy-ion collisions is given by\cite{baurCoherentPhotonphotonInteractions2009}
\begin{equation}
    E_\mathrm{max} = \frac{Ze\gamma}{b^2},
\end{equation}
where Z is the charge of the ion moving along a straight line trajectory, $\gamma$ is the Lorentz boost factor, and $b$ is the distance from the ion's center. For values appropriate at RHIC ($Z=79$, $\gamma\approx100$, $b=15$ fm) one finds a maximum field strength of $E_\mathrm{max}=4.9 \times 10^{16}$ V/cm. Similarly, for values appropriate at the LHC ($Z=82$, $\gamma\approx3000$, $b=15$ fm) an even stronger field of $E_\mathrm{max}=1.5\times10^{18}$ V/cm is found. 

While these fields are well above the critical field strength for Schwinger pair production, they are far from constant --- acting over a short timescale of approximately $\Delta t \simeq b/(\gamma v)$. With $v\sim c$ this gives $\Delta t\simeq10^{-23} (10^{-25})$s at RHIC (LHC). These fields vary far too rapidly to be considered constant with respect to electrons (positrons) (with a Compton wavelength of $\lambda_{C} = 386$ fm~\cite{henckenImpactparameterDependenceTotal1995}). Instead the process must be considered in terms of light quanta. The production of lepton anti-lepton pairs from light quanta has been studied for nearly a century, originating with the seminal works of Breit and Wheeler\cite{breitCollisionTwoLight1934} and Landau and Lifshitz\cite{landauCreationElectronsPositrons1934}, both in 1934.

\subsection{Equivalent Photon Flux}

\begin{figure*}
    \centering
    \begin{subfigure}{.60\textwidth}
      \centering
      \includegraphics[width=.99\linewidth]{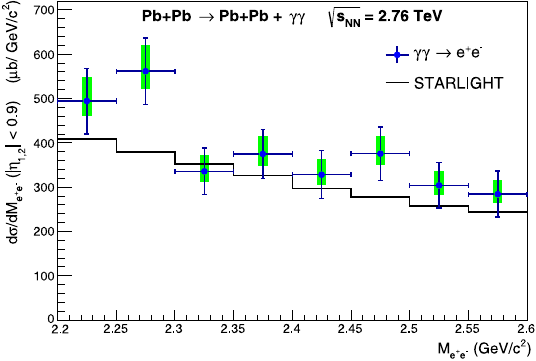}
      \caption{}
    \end{subfigure}%
    \begin{subfigure}{.40\textwidth}
      \centering
      \includegraphics[width=.99\linewidth]{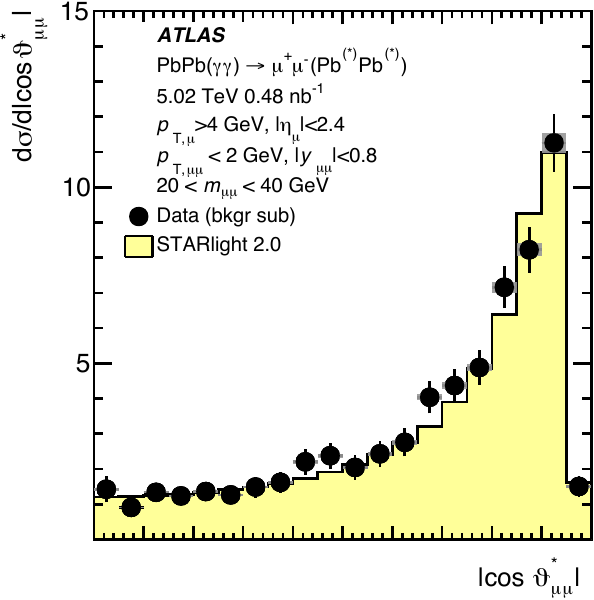}
      \caption{}
    \end{subfigure}
    \caption{
        (color online) 
    \textbf{(a)} The measured $\gamma\gamma \rightarrow e^+e^-$ cross section (blue circles) for ultra-peripheral Pb-Pb collisions in $\sqrt{s_{NN}} = 2.76$ TeV for $-0.9 < \eta < 0.9$ for events in the invariant mass interval $2.2 <M_{\rm inv} < 2.6$ GeV/$c^2$ compared to STARlight~\cite{kleinSTARlightMonteCarlo2017b} simulation (black line). The blue (green) bars show the statistical (systematic) errors, respectively. Reproduced from Ref.~\cite{abbasCharmoniumPairPhotoproduction2013a}.
    \textbf{(b)} 
    The differential cross section for the $\gamma\gamma \rightarrow \mu^+\mu^-$ process shown as a function of $|\cos \vartheta^{\star}_{\mu\mu}|$ in bins of invariant mass ($m_{\mu\mu}$) compared with the cross section prediction from STARlight~\cite{kleinSTARlightMonteCarlo2017b} for $|y_{\mu\mu}| < 0.8$. Statistical uncertainties are shown in error bars while total systematic uncertainties are shown as gray bands. Reproduced from Ref.~\cite{atlascollaborationExclusiveDimuonProduction2020}.
    }
    \label{fig:starlight_data}
\end{figure*}


\label{sec:flux}
Traditionally, in heavy-ion collisions the lowest order pair production process, the $\gamma\gamma \rightarrow l^+l^-$ process (where $l=e, \mu, \tau$), has been studied via the method of equivalent photons, developed by Fermi~\cite{fermiUberTheorieStosses1924}, Weizsacker~\cite{weizsaeckerAusstrahlungBeiStoessen1934}, and Williams~\cite{williamsNatureHighEnergy1934}. According to this method, commonly referred to as the equivalent photon approximation (EPA), the two photon interaction can be factored into a semi-classical and a quantum component. The equivalent flux of photons is dealt with in terms of an external classical field, while the quantum part of the calculation deals with the elementary cross section for the Breit-Wheeler process (in the case of quasi-real photons). 
The cross section for the polarization averaged two-photon process in heavy-ion collisions is then given by
\begin{align}
    \begin{split}
    & \sigma_{\mathrm{A + A}\rightarrow\mathrm{A + A} + l^+l^-}  = \\ & \int  d\omega_1 d\omega_2  n_1(\omega_1) n_2(\omega_2) \times \sigma_{\gamma\gamma\rightarrow l^+l^-}(W; m), 
    \end{split}
\end{align}
where $W$ is the invariant mass of the produced lepton pair and $n_1(\omega_1)$ and $n_2(\omega_2)$ are the equivalent number of photons with energies $\omega_1$ and $\omega_2$ from the field of nucleus 1 and 2, respectively. When the pair transverse momentum $P_\perp$ is small compared to the lepton pair invariant mass $W$, the photon energies are related to the lepton pair invariant mass and rapidity $y$ as
\begin{equation}
    \omega_{1,2} = \frac{W}{2}e^{\pm y},
\end{equation}
and
\begin{equation}
    y=\frac{1}{2}\text{ln}\frac{\omega_{1}}{\omega_{2}}.
\end{equation}
In the external classical field approximation, the heavy nuclei are assumed to travel at constant velocities in straight-line trajectories. 
The polarization-averaged fundamental cross section for the fusion of two real photons into an $l^+l^-$ pair with invariant mass $W$ and lepton mass $m$ is given by the Breit-Wheeler formula~\cite{breitCollisionTwoLight1934,brodskyTwoPhotonMechanismParticle1971a}
\begin{align}
    \begin{split}
    \label{eq:bw}
        \sigma & (\gamma \gamma \rightarrow l^{+}l^{-}) =  \frac{4\pi \alpha_{em}^{2}}{W^{2}} \left[ \left(2+\frac{8m^{2}}{W^{2}} - \frac{16m^{4}}{W^{4}}\right) \right. \\
        & \times\text{ln}\left(\frac{W+\sqrt{W^{2}-4m^{2}}}{2m}\right) \left. -\sqrt{1-\frac{4m^{2}}{W^{2}}}\left(1+\frac{4m^{2}}{W^{2}}\right)\right].
  \end{split}
\end{align}
The number of equivalent photons is determined by equating the energy flux of the transverse electromagnetic field of the moving highly charged ion through an infinitesimal transverse plane element with the energy flux of a photon bunch through the same transverse plane element. For a plane wave, the time-averaged energy flux of the field, described by the Poynting vector, is proportional to the square of the Fourier transformed electric field~\cite{jacksonClassicalElectrodynamics1975}. Therefore the distribution of equivalent photons from the field of one nucleus reads 
\begin{align}
    \begin{split}
    n(\omega; b_\perp) & = \frac{1}{\pi\omega}|E_\perp(b_\perp, \omega)|^2 \\
    & = \frac{4Z^2\alpha_{em}}{\omega} \\
    & \times \left| \int \frac{d^2k_\perp}{(2\pi)^2} k_\perp \frac{F(k^2_\perp +\omega^2/\gamma^2)}{k^2_\perp +\omega^2/\gamma^2} e^{-i b_\perp \cdot k_\perp}  \right|^2,
    \end{split}
    \label{eq:photon-flux}
\end{align}
where $F(k^2)$ is the charge form factor of the nucleus and the photon four-momentum ($k$) is $k = (\omega, k_\perp, \omega/v )$, where $\omega$ is the photon energy, $\gamma$ is the Lorentz-boost factor, and $v$ is the velocity of the nucleus.
In Eq.\ref{eq:photon-flux}, the inherent connection between the total $\gamma\gamma \rightarrow l^+l^-$ cross section, the charge distribution, and the electromagnetic field generated by an ultra-relativistic charged nucleus is evident.

The nuclear charge form factor is computed via the Fourier transform of the charge distribution as
\begin{equation}
    F(k^2) = \int d^3r e^{ ik \cdot r } \rho_{A}( r ).
\end{equation}
The two-parameter Fermi distribution, also known as the Woods-Saxon distribution~\cite{woodsDiffuseSurfaceOptical1954}, is commonly used to describe the charge distribution of high-Z nuclei
\begin{equation}
    \label{eq:woods-saxon}
            \rho_{A}(r)=\frac{\rho^{0}}{1+\exp[(r-R_{\rm{WS}})/d]},
 \end{equation}
where the radius $R_{\rm{WS}}$ (Au: 6.42 fm, Pb: 6.66 fm ) and skin depth $d$ (Au: 0.41 fm, Pb: 0.45 fm~\cite{shouParameterizationDeformedNuclei2015}) are based on fits to electron scattering data~\cite{barrettNuclearSizesStructure1977,devriesNuclearChargedensitydistributionParameters1987a}, and $\rho^{0}$ is the normalization factor. In general, the Fourier transform of the Woods-Saxon distribution must be computed numerically, since it does not have an analytic form. For this reason, many calculations instead prefer to approximate the Woods-Saxon distribution with a hard sphere, with radius $R_A$, convolved with a Yukawa potential with range $a$ (in fm)~\cite{daviesCalculationMomentsPotentials1976}. This charge distribution provides a very good approximation to the Woods-Saxson distribution (See ~\cite{kleinExclusiveVectorMeson1999a} for a direct comparison) and has the analytical form 
\begin{align}
    \begin{split}
    F(|k|) & =  \frac{4\pi\rho^0}{A |k^3|} \left( \frac{1}{1+a^2k^2} \right) \\
    & \times \left[ \sin{(|k| R_A)} - |k| R_A\cos{(|k| R_A)}  \right] , 
    \end{split}
\end{align}
with $A$ as the mass number and where $k$ is the $4-$momentum transfer. Finally, the probability for producing a lepton pair with invariant mass $W$ and rapidity $y$ results from the convolution of the equivalent photon distributions with the elementary cross section, yielding
\begin{align}
    \begin{split}
    P(W,y,b) & = \frac{W}{2}\int d^{2}b_{1\perp}  n(\omega_{1},b_{1\perp}) n(\omega_{2},|\vec{b} - \vec{b}_{1\perp}|) \\ 
    & \times \sigma_{\gamma \gamma \rightarrow l^{+}l^{-}}(W).
    \label{eq:epa_xs}
  \end{split}
\end{align}
 

Traditionally, photon mediated processes, such as the $\gamma\gamma\rightarrow l^+l^-$ interaction, have been studied in ultra-peripheral collisions, where the nuclei pass one another with an ion-ion impact parameter $b$ large enough that there are no hadronic interactions. While this corresponds roughly to the case when $b > 2R_A$, precise treatment of the minimum impact parameter range is influenced by the hadronic overlap probability as a function of impact parameter. Taking this into account, the invariant yield for $\gamma\gamma\rightarrow l^+l^-$ interaction in UPCs can be written as
\begin{equation}
  \frac{d^{2}N}{dWdy}=\int d^{2}b P(W,y,b) \times P_{0H}(\vec{b}),
  \label{equation16}
\end{equation}
where $P_{0H}(b)$ is the probability of not having hadronic interactions, which tends to unity for $b\gg2R_A$. The probability to have no hadronic interaction is 
\begin{equation}
    P_{0H}(b) = 1 - P_{H}(b).
\end{equation}
The hadronic interaction probability ($P_{H}(b)$) is given by
\begin{equation}
  P_{H}(\vec{b}) = 1 - \text{exp}\left[-\sigma_{\text{NN}} \int d^{2}r T_{A}(\vec{r})T_{A}(\vec{r}-\vec{b})\right],
  \label{equation9}
\end{equation}
where $\sigma_{\text{NN}}$ is the inelastic hadronic interaction cross section and $T_{A}(\vec{r})$ is the nuclear thickness function defined as
\begin{equation}
    T_{A}(\vec{r}) = \int dz \rho(\vec{r},z)dz.
  \label{equation10}
\end{equation}
In hadronic collisions, the invariant yield in a given impact parameter range is is given by
\begin{equation}
  \frac{d^{2}N}{dWdy}=\frac{\int^{b_{max}}_{b_{min}} d^{2}b P(W,y,b) \times P_{H}(\vec{b})}{\int^{b_{max}}_{b_{min}} d^{2}b P_{H}(\vec{b})},
  \label{equation16}
\end{equation}
where $b_{min}$ and $b_{max}$ are the minimum and maximum of the impact parameter range, respectively. 
Numerical calculations implementing the equivalent photon approximation according to the above framework, e.g. STARlight~\cite{kleinSTARlightMonteCarlo2017b}, have successfully described the measured $\gamma\gamma \rightarrow l^+l^-$ cross section from several experiments over many years to about the $20\%$ level\cite{abbasCharmoniumPairPhotoproduction2013a,starcollaborationProductionEnsuremathPairs2004,atlascollaborationExclusiveDimuonProduction2020}. 
Figure~\ref{fig:starlight_data} shows comparison between measurements of the $\gamma\gamma$ processes at LHC and STARlight predictions. 
Figure~\ref{fig:starlight_data}a shows measurements of the $\gamma\gamma \rightarrow e^+e^-$ process by the ALICE collaboration as a function of the pair invariant mass ($W=M_{ee}$) along with the STARlight curve. Figure~\ref{fig:starlight_data}b shows measurements of the $\gamma\gamma \rightarrow \mu^+\mu^-$ process by the ATLAS collaboration compared to STARlight predictions as a function of the cosine of the dimuon scattering angle in the $\gamma\gamma$ rest frame, $|\cos \vartheta^{\star}_{\mu\mu}|$, within the ATLAS acceptance and for $|y_{\mu\mu}|<0.8$. 
In both cases the STARlight model provides a reasonable description of the data both in terms of total cross section and differential shape for the invariant mass and rapidity.

\subsection{Kinematics of the $l^+l^-$ Pair}
\label{sec:kinematics}

In the photon-photon fusion process, the produced lepton pair inherits the four-momentum sum of the colliding photons. Specifically, the pair transverse momentum, $P_\perp$, is the vector sum of the two photon transverse momenta
\begin{equation}
    P_\perp = k_{1\perp} + k_{2\perp},
\end{equation}
where $k_{1\perp}$ and $k_{2\perp}$ are the transverse momenta of the photons from the field of nucleus 1 and 2, respectively.
Since ultra-relativistic charged nuclei produce highly-Lorentz contracted electromagnetic fields, the coherently generated photons have momentum predominantly in the beam direction with a transverse component of order $\omega/\gamma$ where $\omega$ is the photon energy and $\gamma$ is the Lorentz factor of the target and projectile nuclei. For this reason the produced pairs have a small transverse momentum component $(P_\perp \simeq \omega/\gamma)$ and the daughter leptons are nearly back-to-back. The deviation from back-to-back can be quantified in terms of the acoplanarity 
\begin{equation}
    \alpha = 1 - \frac{|\phi^+ - \phi^-|}{\pi},
\end{equation}
where $\phi^+$ and $\phi^-$ are the azimuthal angles (angle in plane transverse to the beam) of the two daughter leptons. It should be noted that $\pi\alpha \simeq P_\perp / W$ in an experimental setup where the sagitta of a particle's trajectory is much larger than the effect of multiple scattering in the detector material and from resolution of the experimental measurement. A proper handling of the photon transverse momenta is crucial for accurately describing the lepton pair momentum and $\alpha$ distributions. In the next sections we discuss the theoretical challenges and approaches to computing the photon $k_\perp$ distributions and the resulting pair distributions. 

The allowed spin states for quasi-real photons and the predominantly longitudinal alignment of the photon momentum result in a distinctive polar angle distribution ($\theta$), the angle of a daughter lepton with respect to the beam). The polar angle distribution of these lepton pairs in the photon-photon center-of-mass frame, computed via LOQED, is given by~\cite{brodskyTwoPhotonMechanismParticle1971a}
\begin{align}
    \begin{split}
  G(\theta; W, m) & = 2 + 4(1-\frac{4m^{2}}{W^{2}})  \\
  & \times \frac{(1-\frac{4m^{2}}{W^{2}})\text{sin}^{2}(\theta)\text{cos}^{2}(\theta)+\frac{4m^{2}}{W^{2}}}{(1-(1-\frac{4m^{2}}{W^{2}})\text{cos}^{2}(\theta))^{2}},
  \label{eq:Gtheta}
  \end{split}
\end{align}
for pair invariant mass $W$ and lepton mass $m$. This $G(\theta; W, m)$ distribution results in a distinctive pair rapidity ($y_{ll}$) distribution which results from the quasi-real photon kinematics. Even though most experimental setups are limited to measurement of particles at large angles with respect to the beam, evidence for the distinctive peaking of the pair production at forward/backward angles predicted by the $G(\theta; W, m)$ distribution can still be clearly observed (see Fig.~\ref{fig:starlight_data}b and Ref.~\cite{starcollaborationMeasurementMomentumAngular2021}).

\subsection{Traditional EPA Calculations}
Traditional EPA calculations, e.g. STARlight or those from Ref.~\cite{klusek-gawendaDileptonRadiationHeavyIon2019} determine the photon $k_\perp$ distribution using the so-called $k_\perp$-factorization method (throughout this review we use the term ``$k_\perp$-factorization'' as defined in Refs. ~\cite{klusek-gawendaDileptonRadiationHeavyIon2019,klusek-gawendaCentralityDependenceDilepton2021}). In this approach the one-photon distribution is integrated over all transverse distances, i.e. $0 < b_\perp < \infty$ to obtain the $k_\perp$ distribution (Eq.~16 of Ref.~\cite{kleinSTARlightMonteCarlo2017b})

\begin{equation}
    \frac{dN}{dk_\perp} = \frac{1}{2\pi^2} \frac{F^2(k_\perp^2 + \omega^2/\gamma^2) k_\perp^3}{(k_\perp^2 + \omega^2/\gamma^2)^2}.
    \label{eq:ktfactorization}
\end{equation}
This approach is motivated by the connection between $k_\perp$ and $b_\perp$ as conjugate variables, related by the uncertainty principle\cite{heisenbergUeberAnschaulichenInhalt1927}. For this reason, one cannot readily compute a $b_\perp$ dependent $k_\perp$ distribution. 
At this point it is useful to note that in the EPA, the ion-ion impact parameter is related to the photon-photon interaction location via $\vec{b} = \vec{b}_{1\perp} - \vec{b}_{2\perp}$ - i.e. assuming that the photons are emitted from the center of each nucleus and interact locally.
Even despite this relationship, $k_{1\perp}, k_{2\perp},$ and $P_\perp$ are not directly conjugate to the ion-ion impact parameter $b$, which may be precisely constrained in some experimental setups. 
By virtue of integrating out the $b_{\perp}$ in the one photon distribution, the $k_\perp$-factorization approach leads to a $P_\perp$ distribution which is naturally independent of the ion-ion impact parameter $b$.
It has been shown that such an approximation of the $k_\perp$ distribution differs from the full calculation by order $\omega/\gamma$ - precisely the same order of magnitude as $k_\perp$ itself~\cite{PhysRevC.47.2308,henckenImpactparameterDependenceTotal1995}.
Despite this, the $k_\perp$-factorization approach has achieved some limited success in approximately describing the measured $P_\perp$ and $\alpha$ distributions from various measurements over the last $\sim20$ years~\cite{starcollaborationProductionEnsuremathPairs2004,atlas_collaboration_evidence_2017,atlas_collaboration_observation_2019,atlascollaborationExclusiveDimuonProduction2020,abbasCharmoniumPairPhotoproduction2013a,alicecollaborationMeasurementExcessYield2016b}. 

The first UPC measurement of $\gamma\gamma \rightarrow e^+e^-$ by STAR in 2004~\cite{starcollaborationProductionEnsuremathPairs2004}, with only 52 pairs reconstructed, was the first hint that this impact parameter independent $P_\perp$ distribution could not fully describe the data.
At that time the disagreement between this $k_\perp$-factorization approach and the full lowest-order QED calculations was attributed to significant virtuality which altered the dilepton kinematics~\cite{baltzTwophotonInteractionsNuclear2009,starcollaborationProductionEnsuremathPairs2004}. 
Indeed, in several places over a few decades it has been reported that the quasi-real photons can be treated as real in all cases except with respect to $e^+e^-$ production~\cite{bertulaniPhysicsUltraperipheralNuclear2005a,henckenImpactparameterDependenceTotal1995}. 
However, those conclusions are largely based on the assumption that the disagreement between the EPA calculation and the full lowest-order QED result are due to neglecting the photon virtuality in the EPA case. 
However, as discussed in the introduction, extensive comparisons of various models and LOQED calculations have shown that in the kinematics regimes relevant for production of mid-rapidity pairs with large invariant mass at RHIC and LHC energies, the process and all physical observables are consistent with real photon collisions to high precision (corrections due to virtuality are suppressed by powers of the Lorentz factor, order of $1/\gamma^2)$~\cite{PhysRevC.47.2308}.
Instead the breakdown of the EPA may be due to its inability to describe the changing photon $k_\perp$ distribution as a function of ion-ion impact parameter, and not because of significant photon virtuality - even in the case of $e^+e^-$ pair production.
In the next section we discuss several notable experimental and theoretical developments related to the impact parameter dependence of the $\gamma\gamma \rightarrow l^+l^-$ processes.


\section{Impact Parameter Dependence}
\label{sec:impact}
\begin{figure*}
    \centering
    \begin{subfigure}{.50\textwidth}
      \centering
      \includegraphics[width=.99\linewidth]{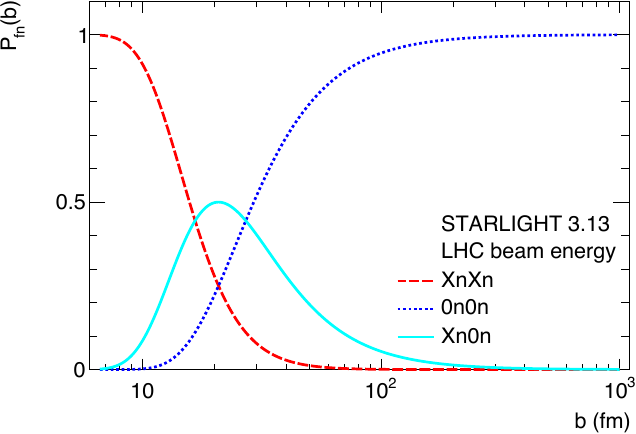}
      \caption{}
    \end{subfigure}%
    \begin{subfigure}{.50\textwidth}
      \centering
      \includegraphics[width=.99\linewidth]{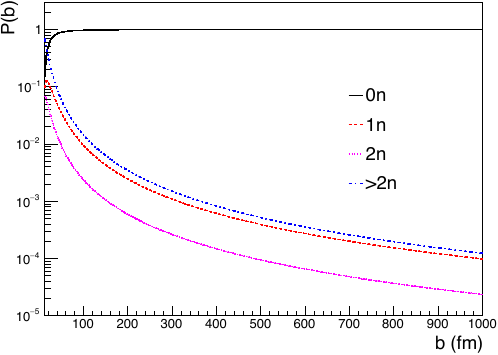}
      \caption{}
    \end{subfigure}
    \caption{ (color online) \textbf{(a)} The impact parameter dependence of the $0$n$0$n, $X$n$0$n, and $X$n$X$n neutron emission scenarios from the STARlight model. Reproduced (with modified line styles) from Ref.~\cite{kleinPhotonuclearTwoPhotonInteractions2020}. \textbf{(b)} Neutron emission probability for $^{208}$Pb as a function of the ion-ion impact parameter in Pb$+$Pb collisons at $\sqrt{s_{\rm{NN}}} = 5.02$ TeV for different neutron multiplicities. Reproduced from Ref.~\cite{brandenburgAcoplanarityQEDPairs2020b}. 
    }
    \label{fig:neutron_theory}
\end{figure*}

\subsection{Neutron Tagging in Ultra-Peripheral Collisions}
Recently, various experimental techniques have been developed to test the impact parameter dependence of the photon flux and kinematics of the produced dileptons. 
For highly charged nuclei, $Z\alpha_{em} \approx 0.6$ ($\alpha_{em} \approx 1/137$ and $Z_{\rm{Au}} = 79$, $Z_{\rm{Pb}} = 82$), meaning that the density of photons is appreciable, and therefore, the nuclei may exchange multiple photons in a single passing.
In UPCs, the quasi-exclusive $\gamma\gamma \rightarrow l^+l^-$ process may be selected in collisions where additional exchanged photons lead to the excitation and subsequent dissociation of the nuclei. 
Mutual Coloumb excitation (MCE) is the process by which at least two photons (in addition to those mediating the semi-exclusive process of interest) cause one or both nuclei to become excited~\cite{stelsonCoulombExcitation1963}. 
The cross section is dominated by the Giant Dipole Resonance (GDR)~\cite{newtonObservationGiantDipole1981} which peaks at low energy ($E_\gamma\approx 14$ MeV/$c$ for gold and lead nuclei). 
The GDR excitation is responsible for several final states with one or two neutrons emitted and has been measured with high precision by various experiments~\cite{veyssierePhotoneutronCrossSections1970}.
Regardless of the specific excitation channel, the presence of forward and/or backward neutrons ($X$n$X$n + $X$n$0$n, where the $X$n$Y$n notation denotes events with $X$ neutrons in one direction and $Y$ neutrons in the other direction) provides a convenient experimental trigger for selecting UPC events that may contain the quasi-exclusive $\gamma\gamma$ processes of interest. 
For this reason, many experiment are well equipped to detect such beam rapidity neutrons via zero-degree calorimeters~\cite{adlerRHICZeroDegree2001b,arnaldiNeutronZeroDegree2006,grachovStatusZeroDegree2006,whiteATLASZeroDegree2010,golubevaNuclearnuclearCollisionCentrality2013}.
However, since hadronic collisions also produce substantial activity in the forward/backward directions, selecting UPC events while simultaneously rejecting events with hadronic overlap generally requires additional detectors in order to verify that no other activity is present besides the quasi-exclusive process of interest (i.e. that a rapidity gap is present). Furthermore, some experiments are well equipped to detect UPC events without any coinciding nuclear dissociation ($0$n$0$n events), providing access to the truly exclusive process of interest~\cite{atlascollaborationExclusiveDimuonProduction2020,cmscollaborationObservationForwardNeutron2020a}.  

The STARlight model provides calculations of the total $\gamma\gamma \rightarrow l^+l^-$ cross section for several common breakup modes including $0$n$0$n, $1$n$1$n, $X$n$X$n, and $0$n$X$n, for gold and lead nuclei based on parameterizations of experimental measurements~\cite{baltzHeavyionPartialBeam1996}. 
Figure~\ref{fig:neutron_theory}a shows the impact parameter dependence of various forward and backward neutron emission scenarios from the STARlight model while Fig.~\ref{fig:neutron_theory}b shows the neutron emission probability for $^{208}$Pb as a function of the ion-ion impact parameter in Pb + Pb collisons at $\sqrt{s_{\rm{NN}}} = 5.02$ TeV from ~\cite{brandenburgAcoplanarityQEDPairs2020b}. 
STARlight contains several common neutron emission scenarios and implements the corresponding impact-parameter dependence for the total cross section calculation but not for the process kinematics. The SuperChic3 model, another common EPA implementation, does not provide specific calculations for any neutron emission scenarios.

Various dedicated models~\cite{pshenichnovElectromagneticExcitationFragmentation2011,pshenichnovMutualHeavyIon2001a,brozGeneratorForwardNeutrons2020} exist for the sole purpose of computing the full neutron emission spectra and can be used for more detailed calculations. In order to calculate the neutron emission spectra, one can begin by computing the mean number of Coulomb excitations resulting from the cloud of Weizsäcker-Williams photons incident on the target nucleus.
The mean number of Coulomb excitations to any state which emits one or more neutrons is given by~\cite{brozGeneratorForwardNeutrons2020}
\begin{equation}
    \label{eq:prob1n}
    m_{Xn}(b) = \int dk n(\omega,b) \sigma_{\gamma A\rightarrow A^{*}}(\omega),
\end{equation}
where $\sigma_{\gamma A\rightarrow A^{*}}(\omega)$ is the photoexcitation cross section for a photon with energy $\omega$.
The photoexcitation cross section $\sigma_{\gamma A\rightarrow A^{*}}(\omega)$ for a given heavy ion can be determined from experimental measurements~\cite{veyssierePhotoneutronCrossSections1970,LEPRETRE1981237,CARLOS1984573,PhysRevD.5.1640,PhysRevD.7.1362,PhysRevLett.39.737,ARMSTRONG1972445}. 
The number of photons with energy $\omega$ at an impact parameter $b$ can be determined by integrating the $k_\perp$ dependence out of Eq.~\ref{eq:photon-flux}.  
The resulting photon spectra, and therefore the mean number of Coulomb excitations, falls off $\propto 1/b^2$ (See Eq.~6 in Ref.~\cite{brozGeneratorForwardNeutrons2020}). 
Therefore, selecting $\gamma\gamma$ events in conjunction with one or more neutron in the forward and backward directions ($X$n$X$n) biases the events toward smaller impact parameters.
Conversely, events can be selected with a veto on forward and backward neutrons ($0$n$0$n), leading to events with a larger average impact parameter.
For this reason, the selection of various neutron emission scenarios provides an experimental technique for investigating the impact parameter dependence of the $\gamma\gamma$ processes since events with more neutrons correspond to a smaller mean impact parameter while events with fewer neutrons correspond to a larger mean impact parameter.

However, $m_{Xn}$ cannot be treated as the neutron emission probability, since $m_{Xn}$ can exceed unity at high energies and small impact parameters, and may lead to non-physical neutron multiplicities (exceeding the number of neutrons in the nucleus). 
Figure~\ref{fig:neutron_theory}b shows the neutron emission probabilities for Pb$^{208}$ following the formalism developed in Refs.~\cite{pshenichnovMutualHeavyIon2001a,PhysRevC.60.044901} which is used by the RELDIS Monte Carlo model~\cite{pshenichnovElectromagneticExcitationFragmentation2011}.
Following this approach, the probability for absorbing exactly N photons is 
\begin{equation}
    P^{(N)}(b) = \frac{m^{N}_{Xn}(b)}{N!}e^{-m_{Xn}(b)},
\label{eq:probN_photons}
\end{equation}
and the corresponding normalized probability density for absorbing $N$ photons with energies $\omega_{1}$, $\omega_{2}$,...,and $\omega_{N}$ is~\cite{brandenburgAcoplanarityQEDPairs2020b}
\begin{equation}
    p^{(N)}(\omega_{1},\omega_{2},...,\omega_{N},b_\perp) =\frac{\prod_{i=1}^{N}n(\omega_{i},b)\sigma_{\gamma A\rightarrow A^{*}}(\omega_{i})}{m^{N}_{Xn}(b_\perp)}.
    \label{eq:photon_prob}
\end{equation}
From this expression the probability density for the first- and Nth-order electromagnetic dissociation process for the channel with emission of $i$ neutrons is expressed as
\begin{equation}
    P^{(1)}_{i}(b_\perp) =\int d\omega_{1} P^{(1)}(b_\perp) p^{(1)}(\omega_{1},b_\perp)f_{i}(\omega_{1}),
    \label{equation7}
\end{equation}
and
\begin{align}
    \label{eq:nthorder_neutrons}
    \begin{split}
    P^{(N)}_{i}(b_\perp) & = \idotsint d\omega_{1}...d\omega_{N} \\
    &\times P^{(N)}(b) p^{(N)}(\omega_{1},...,\omega_{N},b_\perp) f_{i}(\omega_{1},...,\omega_{N}).
    \end{split}
\end{align}    
Here $f_{i}(\omega_{1})$ and $f_{i}(\omega_{1},...,\omega_{N})$ are the branching ratios for the channel corresponding to $i$ neutrons emitted. Assuming the simultaneous absorption of multiple photons gives 
\begin{equation}
    f_{i}(\omega_{1},...,\omega_{N}) =f_{i}(\sum_{k =1}^{N}\omega_{k}).
\end{equation}
Since the higher order contributions given by Eq.~\ref{eq:nthorder_neutrons} fall off very quickly, in practice it is sufficient to compute only the first few terms. Importantly, this approach predicts a well behaved total neutron emission probability even at high energy and small impact parameters.
Detailed calculations of the neutron emission spectra as a function impact parameter dependence are needed for precise comparison between experiments employing neutron tagging and the theoretical calculations that include impact parameter dependence.

\begin{figure*}
    \centering
    \begin{subfigure}{.43\textwidth}
      \centering
      \includegraphics[width=.99\linewidth]{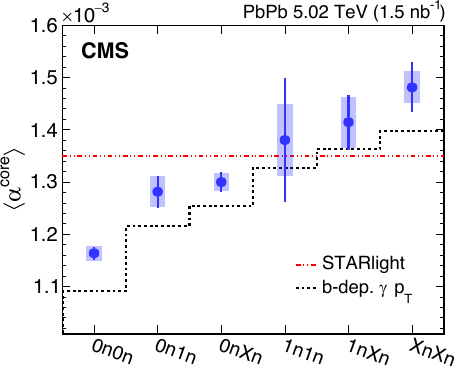}
      \caption{}
    \end{subfigure}%
    \begin{subfigure}{.57\textwidth}
      \centering
      \includegraphics[width=.99\linewidth]{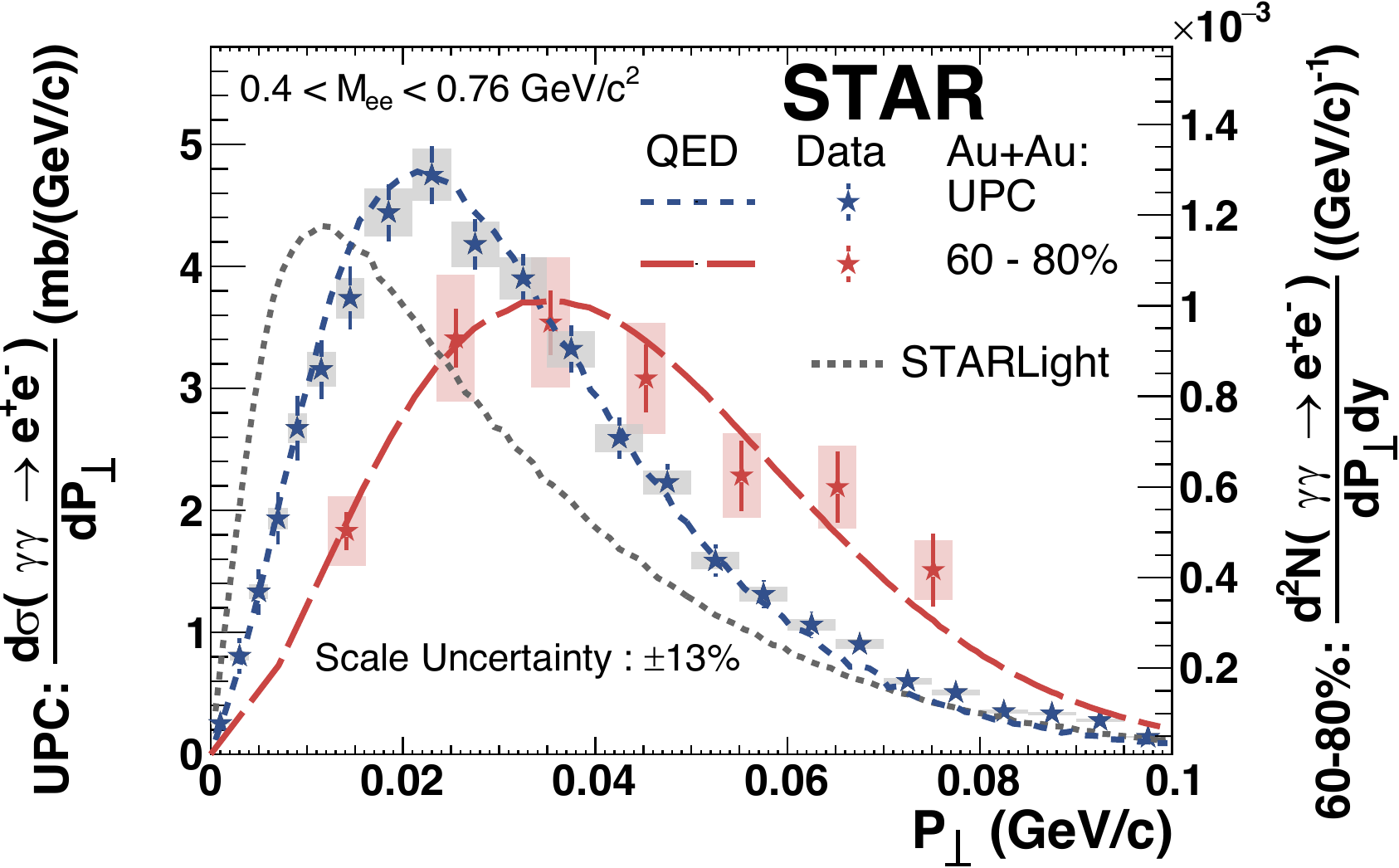}
      \caption{}
    \end{subfigure}
 
    \caption{ (color online) \textbf{(a)} CMS measurement of $\langle \alpha_{\rm core} \rangle$ from the $\gamma\gamma \rightarrow \mu^+\mu^-$ process for various neutron multiplicities compared to the STARlight model and to lowest order QED calculations (b-dep. $\gamma\ p_T$). A strong neutron multiplicity dependence is observed. Reproduced from Ref.~\cite{cmscollaborationObservationForwardNeutron2020a}
    \textbf{(b)} STAR measurement of the $P_\perp$ distribution for the $\gamma\gamma \rightarrow e^+e^-$ process in UPC and $60-80\%$ central hadronic interactions. The measured $P_\perp$ distributions are incompatible with the $k_\perp$-factorization result. Reproduced from Ref.~\cite{starcollaborationMeasurementMomentumAngular2021}.
    }
    \label{fig:exp_impact}
\end{figure*}

The STAR and CMS collaborations have recently employed this neutron tagging approach to experimentally test the impact parameter dependence of the Breit-Wheeler process and to specifically investigate the photon $k_\perp$ distributions.
Figure~\ref{fig:exp_impact}a shows recent measurements by the CMS collaboration of the $\alpha$ distribution ($\pi\alpha \simeq P_\perp / W$) of dimuons in events with various neutron emission scenarios.
A significant dependence of the $\langle\alpha_{\rm core}\rangle$ distribution with neutron multiplicity is observed, where $\alpha_{\rm core}$ is the statistically isolated $\alpha$ distribution from coherent $\gamma\gamma \rightarrow \mu^+\mu^-$ interactions~\cite{cmscollaborationObservationForwardNeutron2020a}. 
In the CMS measurement the narrow signal ($\alpha_{\rm core}$) and broad background distributions were isolated via empirical fit functions. Likewise, STAR has measured the $\gamma\gamma \rightarrow e^+e^-$ process in UPC for events with approximately one neutron in the forward and backward directions ($1$n$1$n)~\cite{starcollaborationMeasurementMomentumAngular2021} and find that the $P_\perp$ distribution is incompatible with the $P_\perp$ distributions predicted by the $k_\perp$-factorization approach (See Fig.\ref{fig:exp_impact}b).
The ATLAS Collaboration has carried out similar measurements of $\alpha$ in the $\gamma\gamma \rightarrow \mu^+\mu^-$ process for events with various numbers of beam energy neutrons~\cite{atlascollaborationExclusiveDimuonProduction2020}. ATLAS found that the total $\alpha$ distribution is generally well describe by STARlight, which models the leading order process, with P\textsc{ythia}8~\cite{sjostrandIntroductionPYTHIA2015} for modeling final-state QED radiation processes and LPair 4.0~\cite{vermaserenTwophotonProcessesVery1983} which models dissociative $\gamma\gamma$ interactions in proton-proton collisions. While ATLAS finds generally good agreement between STARlight and the measured pair rapidity, invariant mass, and $\alpha$ distributions, they do note systematic differences, especially with STARlight's description of the minimum ($k_{\rm min}$) and maximum photon ($k_{\rm max}$) energies~\cite{atlascollaborationExclusiveDimuonProduction2020}. The ATLAS Collaboration further suggests that this may indicate deficiencies in the modeling of the incoming photon flux and could be remedied by adjusting the cutoff radius in the STARlight model~\cite{atlascollaborationExclusiveDimuonProduction2020}.        

Both the STAR~\cite{starcollaborationMeasurementMomentumAngular2021} and CMS~\cite{cmscollaborationObservationForwardNeutron2020a} measurements demonstrate clear and significant impact parameter dependence of the $P_\perp$ (STAR) and $\alpha$ (CMS) distributions for the $\gamma\gamma$ process, and are incompatible with the $k_\perp$-factorization approach.
In addition to the observed impact parameter dependence of the $P_\perp$ and $\alpha$ distributions, both the CMS~\cite{cmscollaborationObservationForwardNeutron2020a} and ATLAS~\cite{atlascollaborationExclusiveDimuonProduction2020} experiments have observed similar modification of dimuon pair invariant mass and rapidity with respect to various neutron emission scenarios. The CMS measurement~\cite{cmscollaborationObservationForwardNeutron2020a} found strong ($>5\sigma$) modification of the pair invariant mass with respect to various neutron emission scenarios, demonstrating a clear impact parameter dependence of the photon energy. While STARlight describes the observed overall trend, it fails to precisely describe the measured invariant mass distributions. Similarly, ATLAS found that STARlight adequately describes the photon energy, except for the minimum ($k_{\rm min}$) and maximum ($k_{\rm max}$) energy photons, where large deviations from STARlight predictions are observed ($\sim20\%$ for $k_{\rm min}$ and $\sim40\%$ for $k_{\rm max}$)~\cite{atlascollaborationExclusiveDimuonProduction2020}. ATLAS measured the dimuon pair rapidity in single dissociative events ($X$n$0$n) and in double dissociative events ($X$n$X$n). While they observe a clear difference between the two sets of events, they find good overall agreement with STARlight within the experimental and theoretical uncertainties. 

In the past it was widely held that the photon virtuality significantly altered the process kinematics (especially for the $e^+e^-$ case) dominating the pair transverse momentum~\cite{baltzPhysicsUltraperipheralCollisions2008}, which would not lead to such impact parameter dependence.
Instead, the observed impact parameter dependence can be understood in terms of a dependence of the photon $k_\perp$ distribution on the EM field distribution, as described by various phenomenological and theoretical approaches discussed in sections~\ref{sec:theory}. 
Crucially, both of these measurements verify the impact parameter dependence of the photon kinematics in the absence of a medium and provide a precise test of the baseline broadening effects due to the electromagnetic field configuration. 
This observation of impact parameter dependence is also important for identifying this interaction as the Breit-Wheeler process~\cite{starcollaborationMeasurementMomentumAngular2021} (as opposed to other processes producing lepton pairs from the fusion of virtual photons~\cite{Landau1934,bethe_h_stopping_1934}) since these measurements identify the source of the $P_\perp$ broadening as originating from the EM field configuration, and not as a result of significant photon virtuality.

\subsection{$\gamma\gamma$ Processes in Hadronic Heavy-Ion Collisions}
Recently the $\gamma\gamma$ processes, have also been identified in peripheral and semi-central events~\cite{starcollaborationLowEnsuremathPair2018b,alicecollaborationMeasurementExcessYield2016b,atlascollaborationObservationCentralityDependentAcoplanarity2018a}. 
Such exclusive $\gamma\gamma$ processes were traditionally thought to occur only in ultra-peripheral collisions given the strong theory assumptions needed for their explanation, namely that the ions generating the electromagnetic fields maintain a straight-line trajectory, and that the photons result from an external electromagnetic field that is coherent both before and after the collision~\cite{zhaCoherentEnsuremathPsi2018}.
Several experiments have observed photo-nuclear~\cite{alicecollaborationMeasurementExcessYield2016b,starcollaborationObservationExcessPsi2019} and photon-photon~\cite{starcollaborationLowEnsuremathPair2018b,atlascollaborationObservationCentralityDependentAcoplanarity2018a} processes in events with hadronic overlap demonstrating the possibility for studying these photon mediated processes even in peripheral to central heavy-ion collisions. 
Measurement of the $\gamma\gamma$ processes in events with hadronic overlap allow access to much smaller impact parameters than are accessible in UPCs (limited by $b_{\min} \approx 2R_{A} $). 
Importantly, in events with hadronic overlap, the collision geometry and impact parameter can be precisely determined via the Glauber approach\cite{millerGlauberModelingHigh2007b,alicecollaborationCentralityDeterminationPbPb2013}.
Furthermore, in these events the $\gamma\gamma$ processes occurs in conjunction with the hadronic interactions that may produce a quark-gluon plasma.

\begin{figure*}
    \centering
      \includegraphics[width=.99\linewidth]{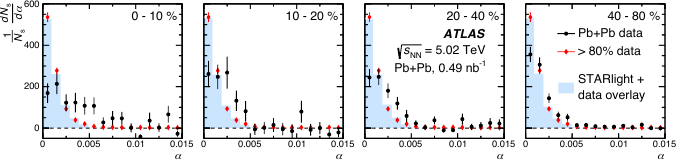}
 
    \caption{ (color online) The background-subtracted distributions (normalized to unity) are shown for $\alpha$. Moving from left to right, the data (circles) are shown for increasingly peripheral collisions. The distributions obtained from STARlight for the $\gamma\gamma \rightarrow \mu^+\mu^-$ process are overlaid on the data and shown for each centrality interval as a filled histogram. The distribution measured in the most peripheral collisions, the $> 80\%$ interval (diamonds) is repeated in each panel to facilitate a direct comparison. The error bars include the statistical and systematic uncertainties. Uncertainties related to the background normalization are not shown. Reproduced from Ref.~\cite{atlascollaborationObservationCentralityDependentAcoplanarity2018a}.
    }
    \label{fig:exp_atlas}
\end{figure*}

Figure~\ref{fig:exp_impact}b shows STAR measurements of $\gamma\gamma \rightarrow e^+e^-$ in UPC and in $60-80\%$ central Au$+$Au collisions where the impact parameter is slightly less than twice the nuclear radius ($b\approx 11.5 - 13.5$ fm). 
The STAR $\gamma\gamma \rightarrow e^+e^-$ result from $60-80\%$ collisions was first compared with the $P_\perp$ distribution from STARlight and found to be significantly broadened~\cite{starcollaborationLowEnsuremathPair2018b}.
In that paper it was proposed that the broadening might result from Lorentz-force bending in the presence of an ultra-strong magnetic field. 
If the pairs were produced in conjunction with a conducting QGP capable of trapping and prolonging the EM fields~\cite{mclerranCommentsElectromagneticField2014}, then the amount of bending would be sensitive to the conductivity of the medium through its effect on the strength and lifetime of the trapped EM fields~\cite{kleinLeptonPairProduction2020a}. 
Similarly, Fig.~\ref{fig:exp_atlas} shows ATLAS measurements of $\alpha$ for the $\gamma\gamma \rightarrow \mu^+\mu^-$ process in peripheral to central Pb$+$Pb collisions that display significant broadening compared to STARlight predictions~\cite{atlascollaborationObservationCentralitydependentAcoplanarity2018}. 
In that paper and others~\cite{mclerranCommentsElectromagneticField2014} it was suggested that broadening may result from electromagnetic multiple scattering analogous to partonic energy loss via strong interactions in a QGP.
In that approach the broadening is quantified in terms of a QED scattering parameter ($\langle\hat{q}_{QED}L\rangle$) of the QGP and the ATLAS $\alpha$ data was found to be well described by $\langle\hat{q}_{QED}L\rangle$ of order ($50$ MeV)$^2$ to ($100$ MeV)$^2$~\cite{kleinLeptonPairProduction2020a,kleinAcoplanarityLeptonPair2019}.
It should be noted that these two mechanisms may not necessarily be mutually exclusive, especially since the STAR and ATLAS measurements cover significantly different kinematic ranges and therefore may probe different dominant mechanisms with the same processes. 
However, both comparisons used baseline distributions from {STARlight}, implementing the $b$-independent $k_\perp$-factorization approach, for determining the amount of medium-induced broadening in $P_\perp$ and $\alpha$.
Clearly, any broadening of the $P_\perp$ and $\alpha$ distributions due to the initial EM field configuration would modify the conclusion about how much, if any, medium induced broadening were present.
As discussed in the previous section, the experimental measurements from STAR~\cite{starcollaborationMeasurementMomentumAngular2021,starcollaborationLowEnsuremathPair2018b} and CMS~\cite{cmscollaborationObservationForwardNeutron2020a} have demonstrated that the photon $k_\perp$-factorization approach lacks impact parameter dependence clearly displayed by the data. 
While the traditional EPA method discussed in Sec.~\ref{sec:process} provides a relatively simple and straightforward technique for computing the $\gamma\gamma \rightarrow l^+l^-$ cross section in heavy-ion collisions, it is insufficient to describe the impact parameter dependence of the photon flux or various effects resulting from the photon polarization. In the next section we review some notable progress made in the phenomenological and theoretical description of the $\gamma\gamma$ processes.

\subsection{Generalized EPA Approach}

\begin{figure*}
    \centering
    \begin{subfigure}{.57\textwidth}
      \centering
      \includegraphics[width=.99\linewidth]{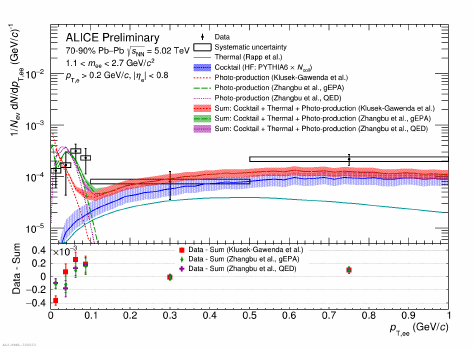}
      \caption{}
    \end{subfigure}%
    \begin{subfigure}{.43\textwidth}
      \centering
      \includegraphics[width=.99\linewidth]{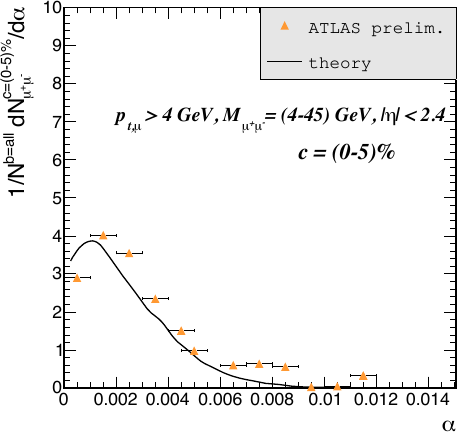}
      \caption{}
    \end{subfigure}
 
    \caption{ (color online) \textbf{(a)} ALICE measurement of $P_\perp$ from the $\gamma\gamma \rightarrow e^+e^-$ process for peripheral Pb$+$Pb collisions~\cite{lehnerDielectronProductionLow2019a}. The data are compared to various models including the EPA $k_\perp$-factorization approach, the gEPA, and the lowest-order QED calculations. Reproduced from Ref.~\cite{lehnerDielectronProductionLow2019a}.
    \textbf{(b)} Comparison between ATLAS measurements of $\alpha$ in central ($0-5\%$) Pb$+$Pb collisions~\cite{atlascollaborationObservationCentralityDependentAcoplanarity2018a} and the predictions from the photon Wigner function formalism~\cite{klusek-gawendaCentralityDependenceDilepton2021}. Reproduced from Ref.~\cite{klusek-gawendaCentralityDependenceDilepton2021}.
    }
    \label{fig:exp_hadronic}
\end{figure*}

\label{sec:theory}
Instead of integrating out the transverse spatial position early on, as is done in the $k_\perp$-factorization method, one can attempt to regain the impact parameter dependence on the pair kinematics by delaying the integration on the interaction location and impact parameter until later. This approach, referred to as a \textit{generalized equivalent photon approximation} (gEPA), was pursued by the authors of Ref.~\cite{zhaInitialTransversemomentumBroadening2020b}. We should emphasize that although the traditional EPA treatment sounds like a mathematical trick to overcome the complication of multiple integrals, and while it remains correct for many observables under the condition without any impact-parameter constraint or tagging, it still has significant consequences. All traditional EPA models ~\cite{kleinSTARlightMonteCarlo2017b,SuperChic3,Zha:2018tlq} assume in their subsequent calculations of impact-parameter dependence that the transverse momentum, spatial location and spin of the photons are not correlated, resulting in impact-parameter independence of many observables~\cite{kleinSTARlightMonteCarlo2017b} and incorrect baseline for experimental results~\cite{starcollaborationLowEnsuremathPair2018b,atlascollaborationObservationCentralityDependentAcoplanarity2018a}, or simply not able to perform such a calculation~\cite{SuperChic3}. 
Again following the procedure of the external classical field approach as in Ref.~\cite{PhysRevC.47.2308}, the gEPA approach starts from the electromagnetic potentials of the two colliding nuclei in the Lorentz gauge:
\begin{equation}
    \label{eq:external}
    \begin{split}
        &A_{1}^{\mu}(k_{1},b_{\tau})= -2 \pi (Z_{1} e) e^{ik_{1}^{\tau}b_{\tau}} \delta(k_{1}^{\nu}u_{1\nu}) \frac{F_{1}(-k_{1}^{\rho}k_{1\rho})}{k_{1}^{\sigma}k_{1\sigma}} u_{1}^{\mu},\\
        &A_{2}^{\mu}(k_{2},b_{\tau}=0)= -2 \pi (Z_{2} e) \delta(k_{2}^{\nu}u_{2\nu}) \frac{F_{2}(-k_{2}^{\rho}k_{2\rho})}{k_{2}^{\sigma}k_{2\sigma}} u_{2}^{\mu}.\\
    \end{split}
\end{equation}
Here $b$ is the impact parameter, which characterizes the separation between the two nuclei. The $\delta$ function ensures that the nuclei travel in straight line motion with a constant velocity. In this approach, the possible deflections from straight line motion due to collisions are neglected, even for events with hadronic overlap. The validity of this approximation will be discussed more in Sec.~\ref{sec:mapping}. The velocities are taken in the center-of-mass frame with $u_{1,2} = \gamma(1,0,0,\pm v)$, where $\gamma$ is the Lorentz contraction factor.

The amplitude for the lepton pair production from the electromagnetic fields in lowest order is then given by the S-matrix element, which leads to Eq.~30 in Ref.~\cite{PhysRevC.47.2308}:
\begin{align}
    \label{eq:gEPA}
    \begin{split}
    \sigma & = 16\frac{Z^{4}e^{4}}{(4\pi)^{2}}\int d^{2}b \int \frac{dw_{1}}{w_{1}} \frac{dw_{2}}{w_{2}} \frac{d^{2}k_{1\bot}}{(2\pi)^{2}} \frac{d^{2}k_{2\bot}}{(2\pi)^{2}} \frac{d^{2}q_{\bot}}{(2\pi)^{2}}\\
     &\times \frac{F(-k_{1}^{2})}{k_{1}^{2}} \frac{F(-k_{2}^{2})}{k_{2}^{2}} \frac{F^{*}(-{k_{1}^{\prime}}^{2})}{{k_{1}^{\prime}}^{2}} \frac{F^{*}(-{k_{2}^{\prime}}^{2})}{{k_{2}^{\prime}}^{2}} e^{-i\vec{b} \cdot \vec{q}_{\bot}} \\
     & \times \big[(\vec{k}_{1\bot} \cdot \vec{k}_{2\bot})(\vec{k}^{\prime}_{1\bot} \cdot \vec{k}^{\prime}_{2\bot}) \sigma_{s}(w_{1},w_{2})\\
     & +(\vec{k}_{1\bot} \times \vec{k}_{2\bot})(\vec{k}^{\prime}_{1\bot} \times \vec{k}^{\prime}_{2\bot})\sigma_{ps}(w_{1},w_{2})\big], 
    \end{split}
\end{align}
where the four momenta of photons are
\begin{align}
    \label{equation4}
    \begin{split}
    k_{1} & = (w_{1}, k_{1\bot},\frac{w_{1}}{v}), k_{2} = (w_{2}, P_{\bot} - k_{1\bot},\frac{w_{2}}{v})\\
    w_{1} & =\frac{1}{2}(P_{0}+vP_{z}), w_{2} = \frac{1}{2}(P_{0}-vP_{z}) \\
    q_{\bot} & = k_{1\bot} -k_{1\bot}^{\prime} \\
    k_{2\bot} & = P_{\bot} - k_{1\bot} \\
    k_{1}^{\prime} & =(w_{1}, k_{1\bot} - q_{\bot},w_{1}/v)\\
    k_{2}^{\prime} & =(w_{2}, k_{2\bot} + q_{\bot},w_{2}/v).\\
    \end{split}
\end{align}
Here the cross section is expressed in terms of the elementary scalar ($\sigma_s$) and pseudo-scalar ($\sigma_{ps}$) cross sections. The scalar and pseudo-scalar cross sections are, as in the case of the total cross section, given by the Breit-Wheeler cross sections for photons with given polarization (either parallel or perpendicular to one another)

\begin{equation}
    \label{eq:bw_pol}
    \begin{aligned}
    \sigma_{\parallel} & = \frac{4\pi \alpha_{em}^{2}}{s} \bigg[\left(2+\frac{8m^{2}}{s} - \frac{24m^{4}}{s^{2}}\right) \\ 
    & \times \text{ln}\left(\frac{\sqrt{s}+\sqrt{s-4m^{2}}}{2m}\right) -\sqrt{1-\frac{4m^{2}}{s}}\left(1+\frac{6m^{2}}{s}\right)\bigg]\\
    \sigma_{\perp} & = \frac{4\pi \alpha_{em}^{2}}{s} \bigg[\left(2+\frac{8m^{2}}{s} - \frac{8m^{4}}{s^{2}}\right) \\
    & \times\text{ln}\left(\frac{\sqrt{s}+\sqrt{s-4m^{2}}}{2m}\right) -\sqrt{1-\frac{4m^{2}}{s}}\left(1+\frac{2m^{2}}{s}\right)\bigg],
    \end{aligned}
\end{equation}
where $m$ is the mass of the produced leptons and $s=4\omega_1\omega_2$. Due to symmetry considerations, the scalar cross section can be identified with the cross section for parallel photon polarization ($\sigma_s=\sigma_\parallel$) and the pseudo-scalar cross section can be identified with the cross section for perpendicular photon polarization ($\sigma_{ps}=\sigma_\perp$). 
Since the total cross section for a fermion pair is a mixture of the scalar and pseudo-scalar components, the total cross section is 
\begin{equation}
    \sigma(\omega_1, \omega_2; m) = \frac{\sigma_{s}(\omega_1, \omega_2; m) + \sigma_{ps}(\omega_1, \omega_2; m)}{2},
\end{equation}
and is given by Eq.~\ref{eq:bw}. It should be noted that an integration of Eq.~\ref{eq:gEPA} gives the well known EPA result. Such a correspondence is a good cross-check and should be required of any impact parameter dependent model. The resulting dilepton kinematic distributions ($P_\perp$, $\alpha$, etc.) can be computed numerical from Eq.~\ref{eq:gEPA}. 
Both the shape and the magnitude of the experimental measurements can be described reasonably well by the gEPA calculations~\cite{zhaInitialTransversemomentumBroadening2020b}. 
Most notably, the $e^{-i\vec{b}\cdot\vec{q_\perp}}$ term found in Eq.~\ref{eq:gEPA} leads to significant impact parameter dependence of the dilepton kinematics ($P_\perp$ and $\alpha$).
However, there is a notable rapid increase in the cross section predicted at low $P_\perp$, which disagrees with the data and also appears to be unphysical. 
The breakdown of the gEPA approach at very low $P_\perp$ may be due to a failure to accurately take into account interference between the two fundamental amplitudes corresponding to the two participating Feynman diagrams for the lowest order process. Put another way, it illustrates an inability to express the process as a convolution of two well defined separate single photon distributions.
In an attempt to remedy this breakdown at low $P_\perp$, the authors of Ref.~\cite{zhaInitialTransversemomentumBroadening2020b} implemented an additional approach by inserting a phase factor of $e^{i\alpha_1 + i\alpha_2}$ into Eq.~\ref{eq:gEPA}, where $\alpha_1$ is the angle between $\vec{k}_{1\bot}$ and $\vec{k}^{\prime}_{1\bot}$ while $\alpha_2$ is the angle between $\vec{k}_{2\bot}$ and $\vec{k}^{\prime}_{2\bot}$.
The additional of this phase factor is phenomenological and motivated by the full lowest-order QED calculations which will be discussed next.

\subsection{Lowest Order QED}
\label{sec:qed}
In lieu of phenomenological models, the full lowest-order (i.e. second order) QED calculations of the $\gamma\gamma \rightarrow l^+l^-$ process can be performed. 
In Refs.~\cite{PhysRevA.51.1874,PhysRevA.55.396} the formalism for the calculation including impact parameter dependence was introduced and results were presented for collisions at an impact parameter of zero ($b=0$ fm) where analytical calculations may be performed. The authors of Ref.~\cite{zhaInitialTransversemomentumBroadening2020b,brandenburgAcoplanarityQEDPairs2020b} extended the calculations to all impact parameters and explored the results relevant for the STAR~\cite{starcollaborationMeasurementMomentumAngular2021,starcollaborationLowEnsuremathPair2018b}, ATLAS~\cite{atlascollaborationObservationCentralityDependentAcoplanarity2018a}, CMS~\cite{cmscollaborationObservationForwardNeutron2020a} and ALICE~\cite{alicecollaborationMeasurementExcessYield2016b} measurements. 

The lowest-order two-photon interaction is a second-order process with two Feynman diagrams contributing, as shown in Fig.~2 of Ref.~\cite{PhysRevA.51.1874,PhysRevA.55.396}. As in the EPA and gEPA, the straight-line approximation for the incoming projectile and target nuclei is applied. 
Following the derivation of Ref.~\cite{PhysRevA.51.1874,PhysRevA.55.396}, the cross section for pair production of leptons is given by
\begin{equation}
\label{equation1_new}
    \sigma = \int d^{2}b \frac{d^{6}P(\vec{b})}{d^{3}p_{+}d^{3}p_{-}} = \int d^{2}q \frac{d^{6}P(\vec{q})}{d^{3}p_{+}d^{3}p_{-}} \int d^{2}b e^{i {\vec{q}} \cdot  {\vec{b}}},
\end{equation}
and the differential probability $\frac{d^{6}P(\vec{q})}{d^{3}p_{+}d^{3}p_{-}}$ in QED at the lowest order is 
\begin{align}
    \label{equation2_new}
    \begin{split}
        \frac{d^{6}P(\vec{q})}{d^{3}p_{+}d^{3}p_{-}} & = (Z\alpha_{em})^{4} \frac{4}{\beta^{2}} \frac{1}{(2\pi)^{6}2\epsilon_{+}2\epsilon_{-}} \\ 
        & \times \int d^{2}q_{1} \frac{F(N_{0})F(N_{1})F(N_{3})F(N_{4})}{N_{0}N_{1}N_{3}N_{4}} \\
        & \times {\rm{Tr}}\bigg\{(\slashed{p}_{-}+m)\bigg[N_{2D}^{-1}\slashed{u}_{1} (\slashed{p}_{-} - \slashed{q}_{1} + m)\slashed{u}_{2} \\
        & + N_{2X}^{-1}\slashed{u}_{2}(\slashed{q}_{1} - \slashed{p}_{+} +m)\slashed{u}_{1}\bigg] (\slashed{p}_{+}-m) \\
        & \times \bigg[N_{5D}^{-1}\slashed{u}_{2} (\slashed{p}_{-} - \slashed{q}_{1} - \slashed{q} + m)\slashed{u}_{1} \\ 
        & + N_{5X}^{-1} \slashed{u}_{1} (\slashed{q}_{1} + \slashed{q} - \slashed{p}_{+} + m)\slashed{u}_{2}\bigg] \bigg\},
    \end{split}
\end{align}
with
\begin{align}
    \label{equation3_new}
    \begin{split}
        N_{0} & = -q_{1}^{2},  \\
        N_{1} & = -[q_{1} - (p_{+}+p_{-})]^{2},\\
        N_{3} & = -(q_{1}+q)^{2}, \\
        N_{4} & = -[q+(q_{1} - p_{+} - p_{-})]^{2}, \\
        N_{2D} & = -(q_{1} - p_{-})^{2} + m^{2},\\
        N_{2X} & = -(q_{1} - p_{+})^{2} + m^{2}, \\
        N_{5D} & = -(q_{1} + q - p_{-})^{2} + m^{2},\\
        N_{5X} & = -(q_{1} + q  - p_{+})^{2} + m^{2},
    \end{split}
\end{align}
where $p_{+}$ and $p_{-}$ are the momenta of the created leptons, the longitudinal components of $q_{1}$ are given by $q_{10} = \frac{1}{2}[(\epsilon_{+} + \epsilon_{-}) + \beta(p_{+z}+p_{-z})]$, $q_{1z} = q_{10}/ \beta$, $\epsilon_{+}$ and $\epsilon_{-}$ are the energies of the produced leptons, and $m$ is the mass of the lepton. In order to compute results at all impact parameters, where in general no simple analytical form is available, the multi-dimensional integration is performed with the VEGAS Monte Carlo integration routine~\cite{peterlepageNewAlgorithmAdaptive1978} in order to compare the predictions directly with the data. 

Like the gEPA result, the lowest order QED calculation predicts strong impact parameter dependence of the dilepton kinematics. The existing measurements from STAR~\cite{starcollaborationLowEnsuremathPair2018b,starcollaborationMeasurementMomentumAngular2021}, ATLAS~\cite{atlascollaborationExclusiveDimuonProduction2020,atlascollaborationObservationCentralityDependentAcoplanarity2018a}, CMS~\cite{cmscollaborationObservationForwardNeutron2020a} and ALICE~\cite{lehnerDielectronProductionLow2019a} all show good agreement, considering experimental uncertainties, with the predictions from the lowest order QED calculations. 
Figure~\ref{fig:exp_impact}b and Fig.~\ref{fig:exp_hadronic}a show comparisons between the $P_\perp$ distribution from lowest order QED calculations and the corresponding measured distributions from data. In both cases the QED calculations show good agreement with the data over several different impact parameter ranges considering the experimental uncertainties.
In addition to the results from Ref.~\cite{zhaInitialTransversemomentumBroadening2020b,brandenburgAcoplanarityQEDPairs2020b}, the authors of Ref.~\cite{liImpactParameterDependence2020} have also carried out full lowest-order QED calculations in order to explore the dilepton transverse momentum and angular distributions in an impact parameter dependent fashion and the two approaches show results that are consistent with one another.
The full implications for mapping of the EM fields based on the results of the QED calculations will be discussed in Sec.~\ref{sec:mapping}.

\subsection{Photon Wigner Distributions}
The Wigner quasiprobability distribution, or just Wigner function, is a method of relating a system's quantum wave function to a quasi-probability function in phase space~\cite{caseWignerFunctionsWeyl2008}. The Wigner function is similar to a classical probability distribution, in that it describes the phase space distribution of the process. However, unlike its classical analog, it does not satisfy all properties of a classical probability function. Specifically, Wigner functions can take on negative values for states of the system that have no classical analogue, a clear indication of quantum interference. The authors of Ref.~\cite{liImpactParameterDependence2020} suggested that the full spatial dependence of the $\gamma\gamma$ processes, including the impact parameter dependence, may be expressed in terms of photon Wigner distributions. 

Since then the authors of Ref.~\cite{kleinLeptonPairProduction2020a,Wang:2021kxm} and Ref.~\cite{klusek-gawendaCentralityDependenceDilepton2021} have investigated the Wigner function formalism for describing the $\gamma\gamma$ processes in heavy-ion collisions.
Following the notation of Ref.~\cite{klusek-gawendaCentralityDependenceDilepton2021}, the photon Wigner function can be expressed as
\begin{align}
  \begin{split}
    \label{eq:Wigner}
    N_{ij} (\omega,b,q) & = \int {d^2 Q \over (2 \pi)^2} \exp[-i b Q] \\ 
    & \times E_i \Big(\omega, q + {Q \over 2} \Big) E^*_j \Big(\omega, q - {Q \over 2} \Big) \, 
    \\
    & = \int d^2s \,  \exp[i q s] \\ 
    & \times E_i \Big(\omega, b + {s \over 2} \Big) E^*_j \Big(\omega, b - {s \over 2} \Big),
  \end{split}
\end{align}
where $E_{i,j}$ are the electric field vectors expressed in terms of the nuclear charge form factors ($F$) as
\begin{eqnarray}
    E (\omega,q) = Z \sqrt{\frac{\alpha_{em}}{\pi}} \,  {q F(q^2+q^2_\parallel) \over q^2 + q^2_\parallel} \,
\end{eqnarray}
where, $q_\parallel = {\omega \over \gamma}$ and $\alpha_{em}\approx 1/137$.

By virtue of the Wigner function definition, Eq.~\ref{eq:Wigner} is a function of both the spatial location ($b$) and transverse momentum ($q$) coordinates.
The standard formula for the photon flux in either momentum space or position space may be obtained by integration over $b$ or $q$, respectively. The differential cross section for lepton pair production can then be expressed in terms of the photon Wigner functions as a convolution over the transverse momenta and transverse positions~\cite{klusek-gawendaCentralityDependenceDilepton2021}
\begin{equation}
    \begin{split}
        \label{eq:mat_space}
        {d\sigma \over d^2b d^2P} &= \int d^2b_1 d^2b_2 \, \delta^{(2)}(b-b_1 + b_2)  \\
        &\times \int {d^2q_1 \over \pi} {d^2q_2 \over \pi} \, \delta^{(2)}(P-q_1 - q_2) \\
        &\times
        \int {d \omega_1 \over \omega_1} {d \omega_2 \over \omega_2} N_{ij} (\omega_1,b_1,q_1) N_{kl} (\omega_2,b_2,q_2)   \\ 
        &\times \, \, {1 \over 2 \hat s} \sum_{\lambda \bar \lambda} M^{\lambda \bar \lambda}_{ik} M^{\lambda \bar \lambda \dagger}_{jl} \, d\Phi(l^+ l^-),
    \end{split}
\end{equation}
where the final term (Eq.~\ref{eq:mat_space}) is a sum over the helicy amplitudes for the $\gamma\gamma \rightarrow l^+l^-$ process and $d\Phi(l^+ l^-)$ is the invariant phase space for the leptons~\cite{klusek-gawendaCentralityDependenceDilepton2021}. The details of the helicity amplitudes are discussed in the next section concerning the photon polarization.

The differential cross section for the $\gamma\gamma \rightarrow \mu^+\mu^-$ process computed via the photon Wigner function formalism is shown in Fig.~\ref{fig:exp_hadronic}b and compared to $\alpha$ distribution data from ATLAS. The Wigner function calculation describes the data reasonably and reproduces the qualitative features of the data, such as the dip in cross section at $\alpha=0$. Indeed, based on the comparisons shown in Ref.~\cite{klusek-gawendaCentralityDependenceDilepton2021,Wang:2021kxm}, it appears that the Wigner function calculations can recover the full impact parameter dependence of the lowest order QED result. 

\section{Photon Polarization}
\label{sec:polarization}

\begin{figure}
    \centering
    \includegraphics[width=0.5\textwidth]{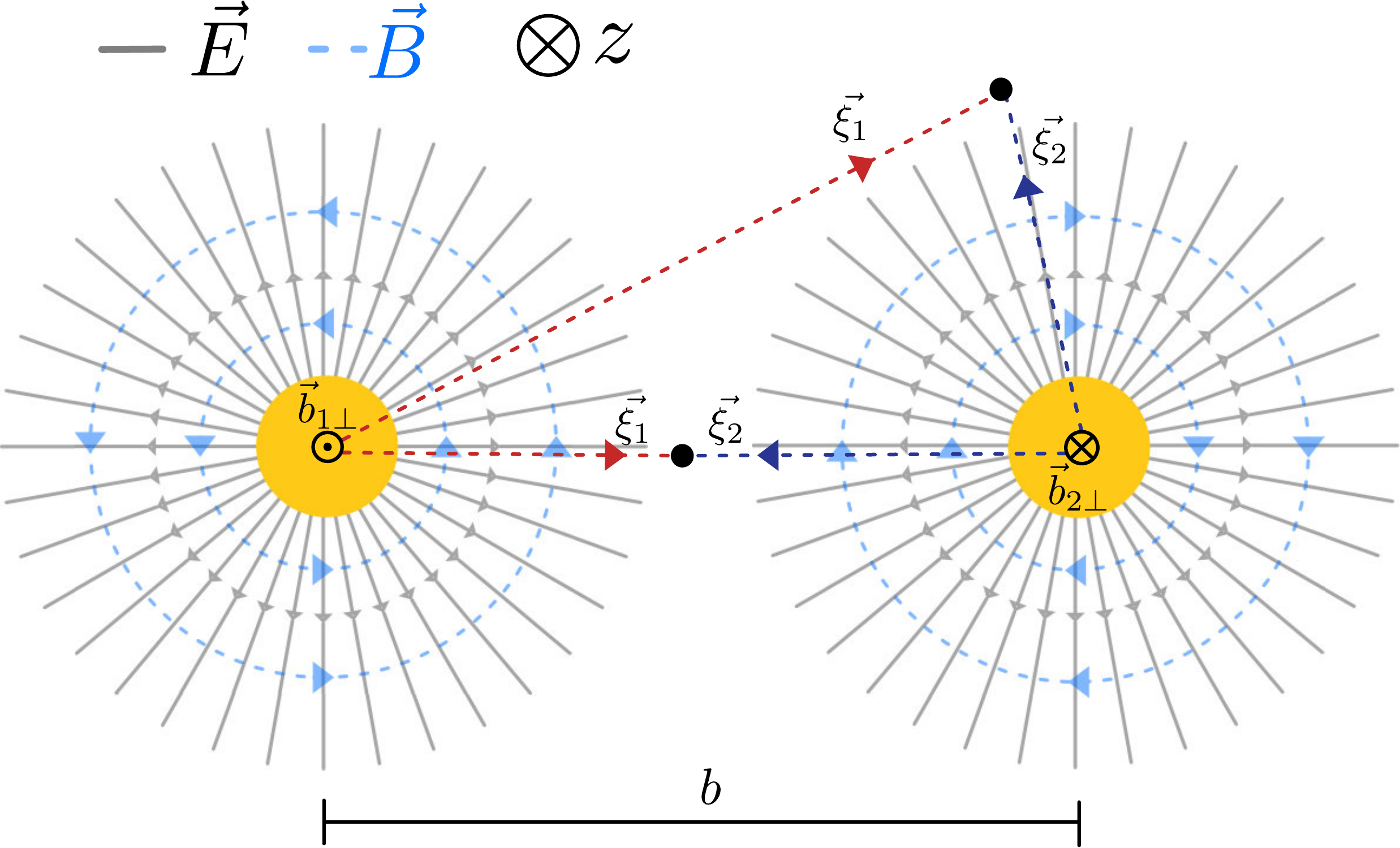}
    \caption{ (color online) Illustration of the electric and magnetic field lines in the plane perpendicular to the beam. The photon interactions locations ($\vec{b}_{1,2\perp}$) are shown for two example interaction locations. Since the dileptons are produced locally, the photon polarization vectors $\vec{\xi}_{1,2}$ are well defined in terms of the classical EM field lines. }
    \label{fig:illustration}
\end{figure}

Since ultra-relativistic nuclei produce a highly Lorentz contracted radial electric field emanating from the nucleus with a magnetic field circling the nucleus, both perpendicular to the direction of motion (See Fig~\ref{fig:illustration}), they propagate as nearly transverse linearly polarized electromagnetic waves at a given infinitesimal transverse plane element. 
Therefore, the coherent photons with vanishing virtuality generated by these fields are expected to be linearly polarized in the transverse plane with respect to the beam. 
However, until recently there was no proposed technique for accessing the photon polarization information, and most theoretical calculations implicitly or explicitly integrated over the polarization (and the resulting azimuthal dependencies)~\cite{SuperChic3}. 
One of the key distinctions between the virtual photon-photon fusion process studied by Landau and Lifshitz~\cite{landauCreationElectronsPositrons1934} compared to the real photon-photon fusion process studied by Breit and Wheeler~\cite{breitCollisionTwoLight1934} is the dependence on photon polarization.
Specifically, Breit and Wheeler predicted different cross sections for photon collisions with parallel vs. perpendicular photon polarization vectors $\vec{\xi}_{1,2}$ (See Eq.~\ref{eq:bw_pol}). 

The authors of \cite{liProbingLinearPolarization2019}, originally applying formalism borrowed from transverse momentum dependent (TMD) parton distribution functions in QCD, recently proposed an experimental signature of linear polarized photon-photon fusion to dileptons.
They predicted that a $\cos4\phi$ modulation is a unique signature of the Breit-Wheeler process with linearly polarized photons, where $\phi$ is the angle between $(p_{1\perp} + p_{2\perp})$ and $(p_{1\perp} - p_{2\perp})/2$ with $p_{1,2\perp}$ being the daughter lepton momenta.  
This result can be understood in terms of the spin and parity of the intermediate two-photon state.

For real photons with helicity $J_z=+,-$ ($J_z=0$ forbidden), the two photon interaction can form states with total helicity and parity($P$) of $J_z^P = 0^\pm, \pm2^\pm$. 
Symmetry properties require that scalar states ($P=+$) form when $E_1 \parallel E_2$ and pseudo-scalar states ($P=-$) when $E_1 \perp E_2$. The $|J_z|=2$ states are dominant when $\beta \rightarrow 1$ while the $J=0$ states dominate for $\beta \ll 0$~\cite{klusek-gawendaCentralityDependenceDilepton2021}, with 
\begin{equation}
    \beta = \sqrt{1 - \frac{4m^2}{W^2}},
\end{equation}
where $\beta$ is the lepton velocity in the dilepton center-of-mass frame, $m$ is the mass of the lepton, and $W$ is the lepton pair invariant mass.
Since the photon polarization is defined by the electric field direction (and perpendicular to the magnetic field) these can be equivalently defined in terms of the photon polarization vectors $\vec{\xi_{1,2}}$.

\begin{figure*}
    \centering
    \begin{subfigure}{.50\textwidth}
      \centering
      \includegraphics[width=0.99\textwidth]{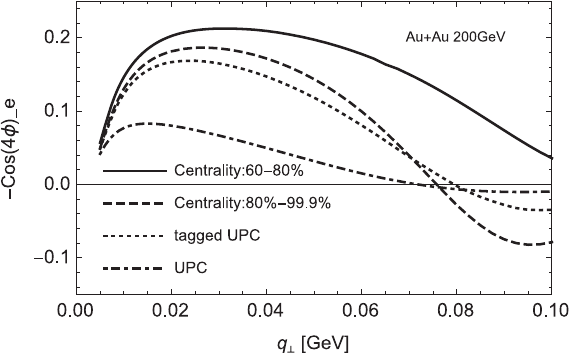}
      \caption{}
    \end{subfigure}%
    \begin{subfigure}{.50\textwidth}
      \centering
      \includegraphics[width=.99\linewidth]{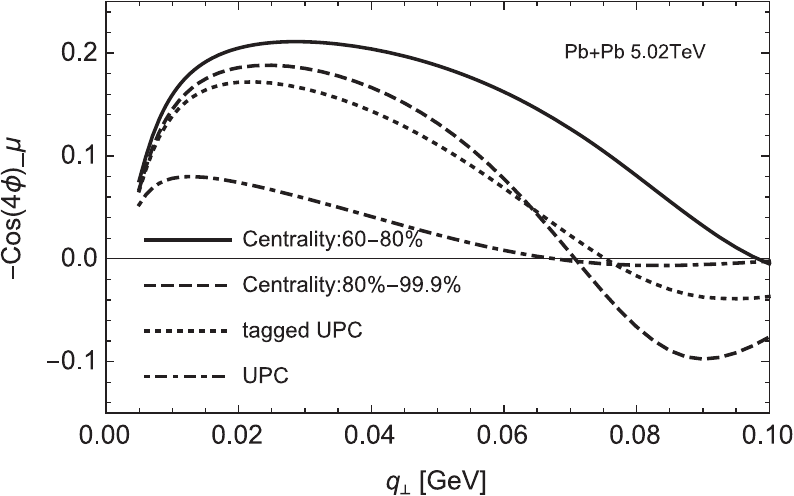}
      \caption{}
    \end{subfigure}
    \caption{(color online) Estimates of the $\cos4\phi$ modulation strength as a function of pair transverse momentum ($q_\perp = P_\perp$) for different impact parameter (centrality) ranges for the $\gamma\gamma \rightarrow e^+e^-$ process in Au$+$Au collisions at $\sqrt{s_{NN}}=200$ GeV \textbf{(a)} and for the $\gamma\gamma \rightarrow \mu^+\mu^-$ process in Pb$+$Pb collisions at $\sqrt{s_{NN}}=5.02$ TeV \textbf{(b)}. The calculations are performed within the STAR and ATLAS acceptances for the $e^+e^-$ and $\mu^+\mu^-$ processes, respectively. Reproduced from Ref.~\cite{liImpactParameterDependence2020}. }
    \label{fig:li_polarization}
\end{figure*}


It is useful to express the Breit-Wheeler process not at the cross section level as before, but at the more fundamental level of the individual helicity amplitudes~\cite{DISPERSIONRELATIONLIGHT,budnev_two-photon_1975,brodsky_two-photon_1971}. 
In the equivalent photon approximation, the $\gamma\gamma \rightarrow l^+l^-$ process at the amplitude level has the form (as taken from SuperChic3~\cite{SuperChic3})

\begin{align}
    \begin{split}
  & k_{1_\perp}^i k_{2_\perp}^j V_{ij} = \\
  & \,
  \begin{cases}     
    -\frac{1}{2} ({\bf k}_{1_\perp}\cdot {\bf k}_{2_\perp})(\mathcal{M}_{++}+\mathcal{M}_{--}) \quad\qquad (J^P_z=0^+)  \\ 
    -\frac{i}{2} |({\bf k}_{1_\perp}\times {\bf k}_{2_\perp})|(\mathcal{M}_{++}-\mathcal{M}_{--}) \qquad  (J^P_z=0^-) \\ 
    +\frac{1}{2}\bigg[(k_{1_\perp}^x k_{2_\perp}^x-k_{1_\perp}^y k_{2_\perp}^y) +i(k_{1_\perp}^x k_{2_\perp}^y+k_{1_\perp}^y k_{2_\perp}^x)\bigg]\mathcal{M}_{-+}  \\ 
    \qquad\qquad\qquad\qquad\qquad\qquad\qquad\qquad\  (J^P_z=+2^+)\\ 
    +\frac{1}{2}\bigg[(k_{1_\perp}^x k_{2_\perp}^x-k_{1_\perp}^y k_{2_\perp}^y) -i(k_{1_\perp}^x k_{2_\perp}^y+k_{1_\perp}^y k_{2_\perp}^x)\bigg]\mathcal{M}_{+-} \\ 
    \qquad\qquad\qquad\qquad\qquad\qquad\qquad\qquad\  (J^P_z=-2^+),     
  \end{cases}\label{Agen}
  \end{split}
\end{align}

where $\mathcal{M}_{\pm \pm}$ corresponds to the $\gamma(\pm) \gamma(\pm) \to X$ helicity amplitude and where $V_{\mu\nu}$ is the $\gamma\gamma \to X$ vertex. 
In this expression it should be noted that the photon transverse momenta $k_{1,2\perp}$ coincide with the photon polarization vectors $\xi_{1,2}$.  This point is crucial for the predicted $\cos4\phi$ modulation in the final state. Without this quantum position-momentum correlation (and therefore spin-momentum correlation) there would be no event-by-event alignment of the photon polarization vectors with the dilepton momentum, and therefore the effects of the photon polarization would average out over many events.

In Ref.~\cite{liProbingLinearPolarization2019} where those authors first predicted the $\cos4\phi$ modulation for polarized photon-photon collisions, they used the impact parameter independent $k_\perp$-factorization approach for computing the photon transverse momentum. 
This approach is equivalent to the implementation of SuperChic3 which uses the $k_\perp$-factorization approach but includes the photon polarization dependent helicity amplitudes in the calculation. 
This approach is not self-consistent since in one case (with respect to the photon transverse momentum) the spatial dependence is integrated out, while on the other hand the spatial dependence is kept when dealing with the polarization. 
The calculation of the $\cos4\phi$ modulation was revisited in Ref.~\cite{liImpactParameterDependence2020} using the full lowest order QED calculation for the collision of real linearly polarized photons in order to take into account the full spatial dependence. 
The resulting $\cos4\phi$ modulation amplitude is shown with respect to the pair transverse momentum ($P_\perp$, denoted $q_\perp$ in the figure) for Au$+$Au (Fig.~\ref{fig:li_polarization}a) and Pb$+$Pb (Fig.~\ref{fig:li_polarization}b) collisions in various impact parameter (centrality) ranges to illustrate the significant dependence on collision geometry. 
While the impact parameter dependent calculation and the $k_\perp$-factorization approach predict a similar average modulation amplitude for $P_\perp \lesssim 0.1 $ GeV/$c$, they predict significantly different $P_\perp$ dependence.
Since the Wigner function formalism maps the system's density matrix, relating the real phase space parameters to the expectation values of Hamiltonian operators, it should be capable of describing the quantum spin-momentum correlations that result in the $\cos4\phi$ modulation of the produced leptons. Indeed, the authors of Ref.~\cite{klusek-gawendaCentralityDependenceDilepton2021} specifically note that the Wigner distribution predicts azimuthal anisotropic behavior and that a future publication will address that aspect of the calculation via the Wigner function in more detail. Ref.\cite{Wang:2021kxm} provided a  general framework for the photon Wigner function and was able to derive the EPA and TMD factorization from this formalism. 

John Toll (Wheeler's student), Euler, and Heisenberg predicted that the vacuum exhibits the optical property of birefringence when subjected to extremely strong electromagnetic fields ($B > B_{C} \approx m_e^2c^3/\hbar e \approx 10^8$ Tesla), the so-called vacuum birefringence effect\cite{DISPERSIONRELATIONLIGHT,heisenbergFolgerungenAusDiracschen1936}. 
Birefringent materials are notable because they exhibit an optical property by which light travels at different speeds (different indices of refraction) depending on whether the polarization plane is parallel or perpendicular to the crystal axis, thus splitting a beam of light into two. 
Vacuum birefringence is a specific example of vacuum polarization in which the path of light splits depending on whether it has a polarization parallel (perpendicular) to the external (magnetic) field direction. 
While vacuum polarization\cite{heisenbergFolgerungenAusDiracschen1936} is responsible for the well-known Lamb shift and Casmir effect\cite{lamb_fine_1947,casimir_influence_1948}, the only reported evidence for vacuum birefringence since its prediction in 1936 has been associated with enhanced linear polarization of light passing through strong magnetic fields near the surface of neutron stars\cite{Mignani:2016fwz} but was debated in the literature on the validity of the conclusion~\cite{Capparelli:2017mlv}. Over the last several decades, several experimental setups, generally employing high powered lasers, have attempted to measure the effect in laboratory settings~\cite{vallePVLASExperimentMeasuring2016}. However, due to the extremely strong fields required and the extreme sensitivity needed to observe the small deflection due to the refraction, laboratory evidence of vacuum birefringence in such experiments has proven unachievable so far. Reference~\cite{PVLASExperiment25} offers an extensive overview of recent efforts to achieve vacuum birefringence via laser-driven experiments.

Vacuum birefringence occurs due to the interaction between the background field(s) and the (virtual) $e^+e^-$ pairs from vacuum fluctuations of the probe photon. 
However, in the case that the $e^+e^-$ masses become real, the probe photon is absorbed instead of transmitted. 
Both the forward scattering and the absorption process can be described by a complex index of refraction.  
This is analogous to classical optics, in which the refraction of light (forward scattering) through an optical medium is described by the real part of the index of refraction, while the  attenuation and absorption is characterized by the imaginary part of the index of refraction~\cite{hechtOptics2015,jacksonClassicalElectrodynamics1975}. Furthermore, the real and imaginary part of the complex index of refraction are related by the Kramers-Kronig relations~\cite{kronigTheoryDispersionXRays1926,DISPERSIONRELATIONLIGHT}.
With vacuum birefringence, even in the absorption process, the imaginary part of the refractive index of empty space depends on the relative angle between the photon polarization and the field orientation, leading to an anisotropic absorption of photons with respect to the magnetic field direction.
In the formalism of QED, the forward scattering process is connected to the absorption process through the optical theorem~\cite{budnev_two-photon_1975}.
Unlike the Schwinger mechanism, which is necessarily a non-perturbative effect, the fundamental effect leading to vacuum birefringence can be identified in the lowest order Feynman diagram, as illustrated in Fig.~1 of Ref.~\cite{vallePVLASExperimentMeasuring2016}. 

The Breit-Wheeler process in heavy-ion collisions can be identified as the absorption of a linearly polarized photon from one nucleus by the intense electromagnetic field with a fixed circular magnetic component around the other nucleus (i.e. another source of linearly polarized photons). 
The resulting angular dependence of the photon absorption rate is therefore related to the lowest order effect contributing to vacuum birefringence\cite{hattoriVacuumBirefringenceStrong2013,Mignani:2016fwz}. 
In fact, in retrospect it can be seen that Toll's thesis, which builds upon the work of Breit and Wheeler, actually predicts the anisotropic effect arising from the photon polarization (in that case without interference, so a $\cos2\phi$ modulation) as can be seen in Eq. 4.3-7 of Ref.~\cite{DISPERSIONRELATIONLIGHT}. In a few other places the relation between the indices of refraction and the lowest order effects in both the forward scattering and pair production process have been explored~\cite{dinuVacuumRefractiveIndices2014,dinuPhotonPolarisationLightbylight2014}.
 
At this time only the STAR experiment has measured the $\cos4\phi$ modulation~\cite{starcollaborationMeasurementMomentumAngular2021} resulting from the anisotropic vacuum polarization effects. The modulation strength and centrality dependence observed in the $\gamma\gamma \rightarrow e^+e^-$ process show good agreement (within $\pm1\sigma$) with the full lowest-order QED calculations for the collision of real photons with linear polarization that take into account impact parameter dependence. The STAR result in $60-80\%$ central collisions display a stronger modulation ($-2\times\langle\cos4\phi\rangle = 27\pm6\%$, including stat. and syst. uncertainty) compared to the same measurement in neutron tagged ultra-peripheral collisions ($-2\times\langle\cos4\phi\rangle = 16.8\pm2.5\%$), providing evidence for the predicted impact parameter dependence.
However, the reported experimental uncertainties are too large to definitively demonstrate impact parameter dependence on the average $\cos4\phi$ modulation strength. Future opportunities at STAR, possible with higher precision multi-differential measurements, will be discussed in Sec.~\ref{sec:future}.

\section{Mapping the Electromagnetic Fields}
\label{sec:mapping}
As discussed in Sec.~\ref{sec:process}, the order-of-magnitude electromagnetic field strengths in heavy-ion collisions can be computed fairly straight-forwardly, while much more involved calculations are needed to predict the detailed spatial distribution~\cite{skokovEstimateMagneticField2009} and time evolution of the fields~\cite{mclerranCommentsElectromagneticField2014,kharzeevChiralMagneticConductivity2009}.
Furthermore, detailed calculations involving event-by-event fluctuation of the charge distribution (protons) within the nucleus indicate that the electric and magnetic field strength, and orientation, may in some cases change dramatically with respect to the case for a static and continuous charge distribution~\cite{bzdakEventbyeventFluctuationsMagnetic2012}. 
However, despite important implications on emergent magnetohydrodynamical phenomena of QCD, as of yet there has been little in the way of experimental measurements capable of constraining the electromagnetic field distributions in their spatial extent or in terms of their time evolution.

The ability to map the electromagnetic field produced in heavy-ion collisions using the $\gamma\gamma$ processes is based on the following specific principles:
\begin{itemize}
  \item The total production cross section, via the photon flux (see e.g. Eq.~\ref{eq:photon-flux}, ~\ref{eq:Wigner}), is indicative of the transverse electric field strength.
  \item The pair transverse momentum distribution is influenced by the collision geometry due to the spatial distribution of the electromagnetic fields. 
  \item Observation of the predicted $\cos4\phi$ modulation experimentally demonstrates the photon polarization states and therefore provides a clear connection between the electric and magnetic field strengths (in a given frame).
\end{itemize}

Each of these aspects will be discussed in more detail in the following sections. Here it is valuable to point out that the in the case of the external field approximation, the electric and magnetic fields (in a given frame) are fully defined by the electromagnetic four-potential (see Eq.~\ref{eq:external}) and that the only free parameters in this description are the photon momenta and the nuclear charge distribution (we assume that the uncertainty on the Lorentz-boost factor $\gamma$ is negligible and that the charge $Z$ is known exactly). Therefore, in the following discussion the mapping of field the distribution is discussed primarily in terms of the ability to constrain the photon kinematics and the nuclear charge form factor. 

\begin{figure*}
  \centering
  \begin{subfigure}{.49\textwidth}
    \centering
    \includegraphics[width=0.99\textwidth]{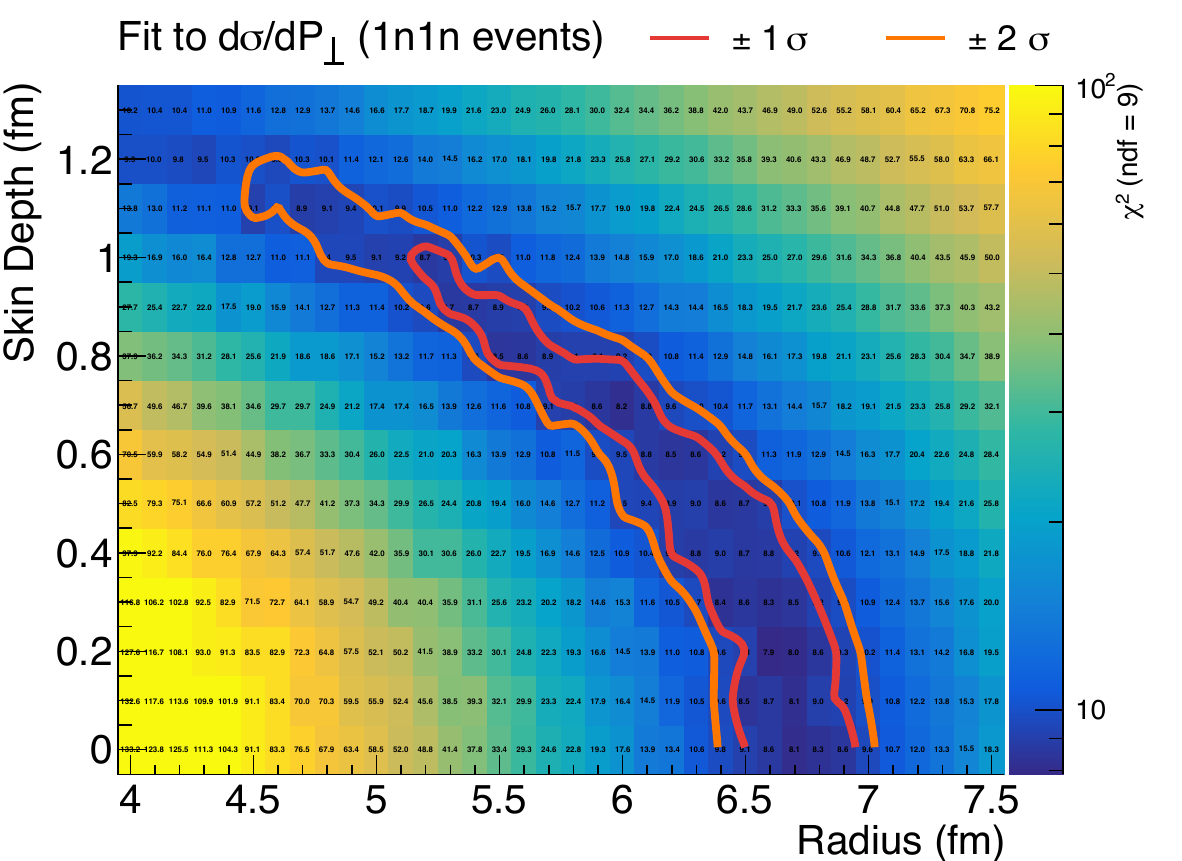}
    \caption{}
  \end{subfigure}%
  \begin{subfigure}{.51\textwidth}
    \centering
    \includegraphics[width=.99\linewidth]{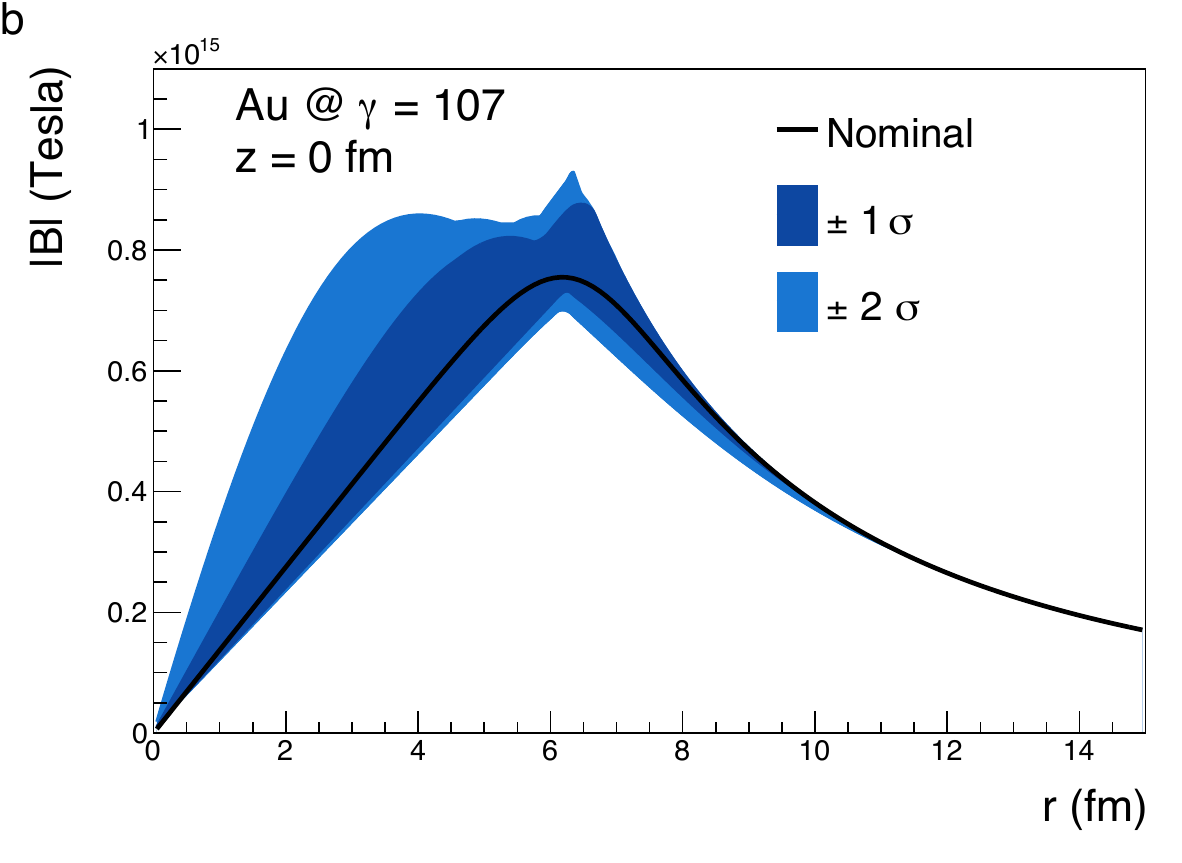}
    \caption{}
  \end{subfigure}
  \caption{(color online) \textbf{(a)} The $\chi^2$ landscape obtained from a fit to the STAR measured $P_\perp$ distribution from the $\gamma\gamma \rightarrow e^+e^-$ process in $1$n$1$n events using the lowest order QED calculations (Sec.~\ref{sec:qed}). Reproduced from Ref.~\cite{starcollaborationMeasurementMomentumAngular2021}. The nuclear charge distribution is assumed to follow a continuous charge distribution given by a Woods-Saxon distribution. \textbf{(b)} The strength of the magnetic field produced by a relativistic heavy-ion (with Lorentz-boost factor $\gamma=107$) determined through a fit to the $P_\perp$ distribution. Reproduced from Ref.~\cite{starcollaborationMeasurementMomentumAngular2021} }
  \label{fig:b_mapping}
\end{figure*}

\subsection{Transverse Momentum and Invariant Mass}
The success of the gEPA, lowest order QED, and Wigner function formalism of the $\gamma\gamma$ process to describe the impact parameter dependence observed by various experiments illustrates the importance of considering the spatial distribution of the electromagnetic fields. 
While the full QED calculations contain all of the relevant physics, they are not as easily interpretable as the gEPA or Wigner function formalism. 
As discussed in Ref.~\cite{kleinLeptonPairProduction2020a}, to first order, the photon $k_\perp$ is related to the transverse interaction distance $b_\perp$ as $\langle k_\perp^2 \rangle \propto 1/b_\perp^2$. 
Since the coherent photon flux from the full charge of the nucleus ($Z$) cannot participate for $b_\perp < R_A$, the characteristic momentum for photons from the field of a given nucleus with charge radius $R_A$ is $\langle k_\perp^2 \rangle \propto 1 / R_A^2$.
Additionally, since the pair transverse momentum is the vector sum of the individual photon momenta ($k_{1\perp}, k_{2\perp}$), one can intuitively see the connection between the charge distribution and the final state pair transverse momentum.

As an example of the constraining power, Figure~\ref{fig:b_mapping}a shows the results of a $\chi^2$-minimization procedure applied by fitting the STAR measured $P_\perp$ distribution from the $\gamma\gamma \rightarrow e^+e^-$ process in $1$n$1$n events using the lowest-order QED calculations (Sec.~\ref{sec:qed})~\cite{starcollaborationMeasurementMomentumAngular2021}. For the minimization, the nuclear charge distribution is parameterized according to a Woods-Saxon with both the radius and skin depth (i.e. surface thickness, not to be confused with neutron skin) as free parameters. 
Figure~\ref{fig:b_mapping}b shows the corresponding magnetic field distribution allowed by the nuclear charge distribution, noting that the connection to the $B-$field strength requires simultaneous measurement of the photon polarization~\cite{starcollaborationMeasurementMomentumAngular2021}, as was carried out by STAR. 
The uncertainty from the data on the measured $P_\perp$ distribution are translated into $\pm1\sigma$ and $\pm2\sigma$ uncertainties on the magnetic field distribution. 
The data are, within uncertainty, fully consistent with the nominal field strength given by the expected distribution for a continuous charge distribution determined by a Woods-Saxon with parameters $R_A=6.38\ {\rm fm}, a=0.535$ fm (black solid line in Fig.~\ref{fig:b_mapping}b). 
The measured distribution and uncertainties do not allow a significantly weaker field (compared to the nominal), but do allow significantly stronger field strengths at small radii, within the nucleus. 
This reduced constraining power for small radial distances is largely due to the limited range of the STAR $P_\perp$ measurement (only for $P_\perp < 0.1$ GeV/$c$), since the field at small distances contributes predominantly to the cross section at higher $P_\perp$.
Additional precision measurements at higher $P_\perp$ will be especially important for constraining the effects of event-by-event fluctuations of the nuclear charge distribution, since these are expected to produce changes in the field distribution that are largest at small radii within the nucleus~\cite{skokovEstimateMagneticField2009}.

\begin{figure}
  \centering
  \includegraphics[width=.4\textwidth]{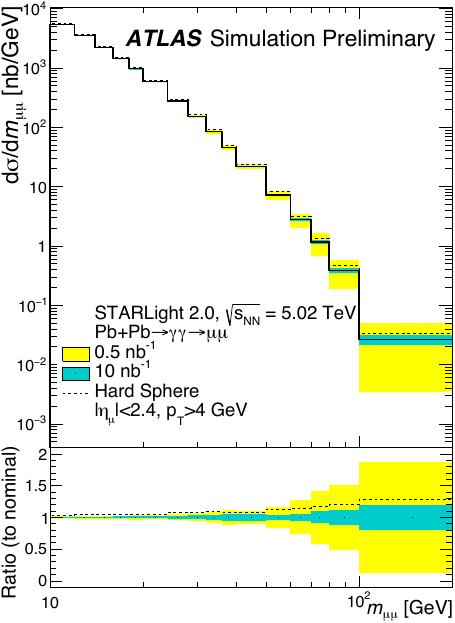}
  \caption{(color online) STARlight~\cite{kleinSTARlightMonteCarlo2017b} predictions for the differential cross section for exclusive production of the dimuon pairs as a function of the invariant mass. Two nuclear charged form factors are considered: a realistic skin depth of the nucleus (solid line) or a hard sphere (dashed line). The lower panel shows the ratio to nominal as a function of the invariant mass, where nominal is the case with the realistic skin depth of the nucleus~\cite{ProspectsMeasurementsPhotonInduced}. Reproduced from Ref.~\cite{ProspectsMeasurementsPhotonInduced}. }
  \label{fig:atlas_sim}
\end{figure}

The $\gamma\gamma$ processes is also sensitive to the details of the nuclear charge form factor through the dilepton invariant mass distribution, since the charge form factor regulates the photon energy. While the dominant behavior is driven by the characteristic mass scaling of the Breit-Wheeler cross section (see Eq.~\ref{eq:bw}), some effect is expected due to the energy dependence of the photon flux which is partially determined by the charge form factor. Specifically, the higher frequency (finer spatial details, e.g. hard edges) of the charge form factor are resolved by higher invariant mass pairs~\cite{klusek-gawendaExclusiveMuonpairProduction2010}. 
As an example of this, Fig.~\ref{fig:atlas_sim} shows an ATLAS simulation of the invariant mass distribution of $\mu^+\mu^-$ pairs for a realistic charge distribution compared to a hard sphere~\cite{ProspectsMeasurementsPhotonInduced}. A measureable difference, considering the predicted uncertainties for future data samples, is visible between the two charge distributions with the most significant difference at higher invariant masses.

The impact parameter dependence of the total $\gamma\gamma \rightarrow l^+l^-$ cross section is also sensitive to the spatial distribution of the electromagnetic fields. With recent measurements of the $\gamma\gamma$ process in events with hadronic overlap, precise tests of the impact parameter dependence are possible for the first time. Observation of the photon mediated processes in events with hadronic overlap also provide an opportunity to test the applicability of the external field approximation. There are two aspects of the approximation that may be ill suited for describing events with hadronic overlap: first, the assumption that the fields are generated by nuclei that are undeflected and travel along straight-line trajectories, and second, that the participating coherent electromagnetic fields are a result of the full charge $Z$ of the colliding nucleus. It has been suggested that in events with hadronic collisions, the external field applicable to the $\gamma\gamma$ process may result only from the portion of the nucleus participating in the collision, or alternatively, only from the portion not-interacting (spectators) since that portion is most likely to follow the assumed undeflected, straight-line trajectory~\cite{zhaCoherentEnsuremathPsi2018,starcollaborationObservationExcessPsi2019,alicecollaborationMeasurementExcessYield2016b}. 

However, at this time, the predictions using the unmodified external field approximation appear to roughly describe the existing measurements within experimental uncertainties. The recent ATLAS measurements of the $\gamma\gamma \rightarrow \mu^+\mu^-$ process over a wide range of centralities are currently best suited to test the applicability of the external field approximation in this way. The $\alpha$ distribution calculated using the Wigner function formalism~\cite{klusek-gawendaCentralityDependenceDilepton2021} compared to the ATLAS measurement~\cite{atlascollaborationObservationCentralityDependentAcoplanarity2018a} in $0-5\%$ most central collisions shows good agreement, both in terms of the differential shape and the absolute normalization (see Fig.~\ref{fig:exp_hadronic}b). The good agreement would suggest that the external field approximation is, at least to leading order, a good approximation even in central heavy-ion collisions. Future high precision and multi-differential measurements will provide deeper insights into the realm of applicability of the external field approximation.

\subsection{Photon Polarization and Collision Geometry}
Compared to the transverse momentum distribution alone, the $P_\perp$ dependence of the $\cos4\phi$ modulation provides even more information for constraining the photon kinematics and charge distribution. As depicted in Fig.~\ref{fig:illustration}, the final amplitude of the $\cos4\phi$ modulation is a balance between positive contributions when $\xi_1 \perp \xi_2$ and negative contributions when $\xi_1 \parallel \xi_2$. 

Pairs with low momentum, where the photons collide with approximately anti-parallel and nearly equal transverse momenta, have polarization vectors that are correspondingly anti-parallel. Since the photon polarization is approximately anti-parallel, the pairs at low $P_\perp$ are expected to display an overall negative $\cos4\phi$ modulation amplitude.
In contrast, pairs with slightly higher transverse momentum are predominantly produced by photons that interact nearly perpendicular, which leads to a predominantly positive $\cos4\phi$ modulation. The QED calculations of the process~\cite{liImpactParameterDependence2020} show this oscillation between an overall negative and positive modulation as a function of the $P_\perp$. Figure~\ref{fig:li_polarization}a and b show the predicted $\cos4\phi$ modulation strength for Au$+$Au (a) and Pb$+$Pb (b) as a function of $P_\perp$ and illustrates the sensitivity to the nuclear charge distribution. 
While the peak modulation amplitude is nearly equal between the two cases, the $P_\perp$ dependence shows a smaller oscillation period for the larger Pb nucleus ($R_{\rm Pb} = 6.62$ fm) compared to Au ($R_{\rm Au} = 6.38$ fm). 
Figure~\ref{fig:li_polarization} also indicates that some impact parameter ranges may be more sensitive to the nuclear charge distribution than others, especially since the cross section for the coherent $\gamma\gamma \rightarrow l^+l^-$ process drops rapidly with $P_\perp$, making measurement above $P_\perp\approx 0.1 {\rm\ GeV}/c$ difficult.

\begin{figure}
    \centering
    \includegraphics[width=.99\linewidth]{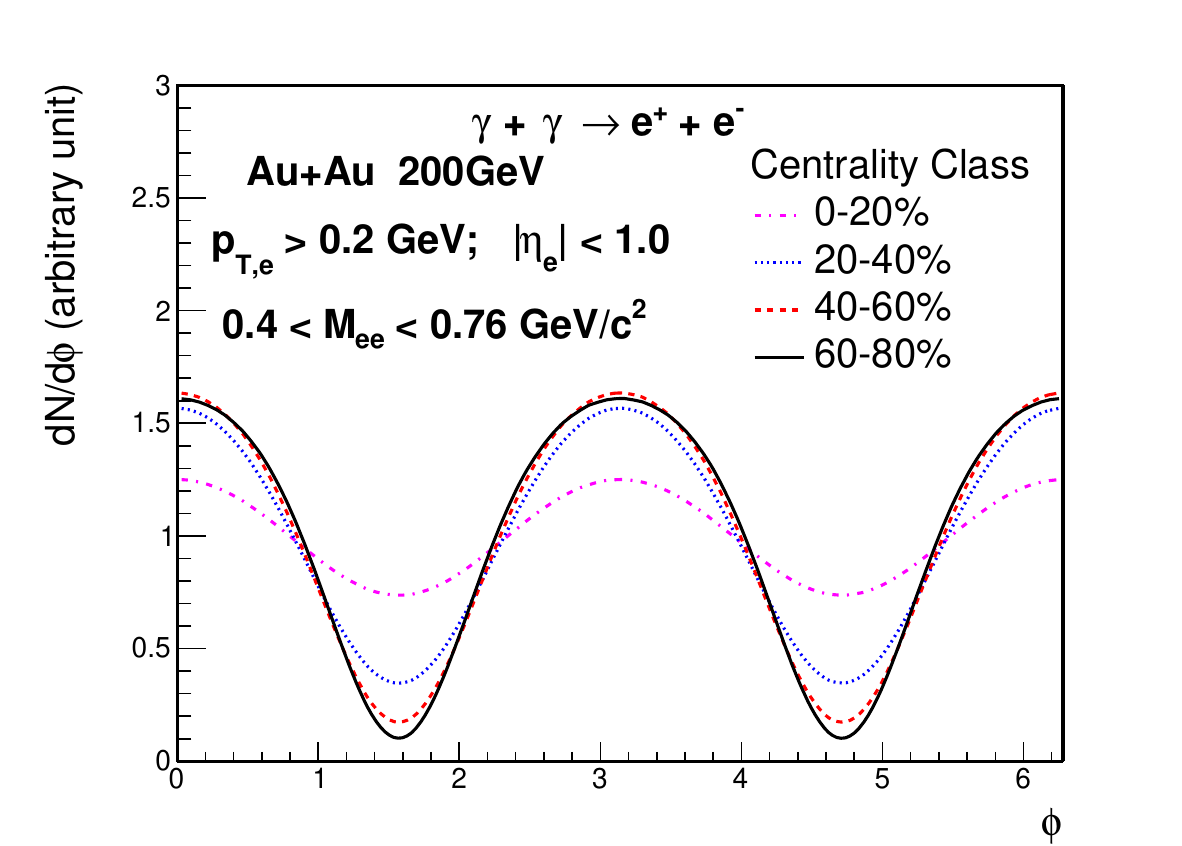}
    \caption{(color online) 
    The azimuthal angular distributions of $e^+e^-$ pairs ($M_{ee}=0.6 {\rm GeV}/c^2$ at y=0) with respect to the reaction plane for different centralities in Au+Au collisions at $\sqrt{s_{NN}} = 200 $ GeV~\cite{zhaInitialTransversemomentumBroadening2020b}. 
  }
  \label{fig:future}
\end{figure}

\subsection{Final State and Medium Effects}
The $\gamma\gamma$ processes that were previously considered viable only in ultra-peripheral collisions have recently been measurements by STAR~\cite{starcollaborationProductionEnsuremathPairs2004} and ATLAS~\cite{atlascollaborationObservationCentralitydependentAcoplanarity2018} in heavy-ion collisions with hadronic overlap. 
The simultaneous measurement of the purely QED-driven $\gamma\gamma$ processes alongside the production of a quark-gluon plasma in hadronic collisions may in principle provide an opportunity to use the QED processes as precision probes of the QCD medium.
A significant broadening effect in $P_\perp$ (STAR) and $\alpha$ (ATLAS) was observed with respect to the same distribution in UPCs and to predictions from EPA calculations.
Since the concept of a broadening in the photon momentum with respect to impact parameter was not considered at that time, various possible mechanisms were offered to explain the significant broadening observed with respect to existing EPA calculations.
The STAR paper~\cite{starcollaborationProductionEnsuremathPairs2004} offered a possible broadening mechanism resulting from Lorentz-force bending as the daughter leptons traverse the strong magnetic field generated in the collision.
On the other hand, the ATLAS result was compared with models that involved multiple scattering as the daughter leptons traversed the produced medium~\cite{atlascollaborationObservationCentralityDependentAcoplanarity2018a}.
In both cases, the fundamental assumption was that the existing EPA calculations were a proper baseline for extracting possible medium induced or final state effects.

As discussed in Sec.~\ref{sec:impact}-\ref{sec:polarization}, the photon kinematics and the produced dilepton kinematics show strong dependence on the collision geometry.
Both the STAR~\cite{starcollaborationProductionEnsuremathPairs2004} and ATLAS~\cite{atlascollaborationObservationCentralityDependentAcoplanarity2018a} data are in relatively good agreement with the theoretical calculations that take into account the spatial dependence of the fields (gEPA~\cite{zhaInitialTransversemomentumBroadening2020b}, QED~\cite{zhaInitialTransversemomentumBroadening2020b,liImpactParameterDependence2020}, and Wigner function formalisms~\cite{klusek-gawendaCentralityDependenceDilepton2021,kleinLeptonPairProduction2020a,Wang:2021kxm}).
Considering the good agreement with these calculations that include the vacuum process only, room for any such medium induced or final state effects is significantly reduced. 
Higher precision measurements will be needed to determine the level at which these additional effects may be present. 
For instance, the existing STAR $P_\perp$ measurement, while consistent with the QED prediction alone, favors a best fit with a slightly broader distribution characterized by an additional $k_\perp$ kick of $14\pm4 {\rm\ (stat.)} \pm4 {\rm\ (syst.)}$ MeV/c .
The ATLAS collaboration has recently completed a new, higher precision measurement of the $\gamma\gamma \rightarrow \mu^+\mu^-$ process~\cite{atlascollaborationExclusiveDimuonProduction2020} which may provide additional power for distinguishing the effects of medium interaction or Lorentz-force bending in a magnetic field. 
Furthermore, it has recently been suggested that the source of any additional broadening may be determined 
through measurement of the broadening with respect to the rapidity difference between the daughter leptons~\cite{kleinLeptonPairProduction2020a}.
Since the Lorentz force applied to a charged particle with velocity $v$ is proportional to $\vec{v} \times \vec{B}$ the leptons with equal rapidity (and approximately equal momentum) will result in cancellation of the magnetic field bending effects. The analysis in Ref.~\cite{kleinLeptonPairProduction2020a} suggests that the magnetic field broadening begins to saturate for a rapidity difference between the two leptons of approximately 3 units. Therefore, differential measurements of the transverse momentum spectra with respect to the rapidity difference may provide more quantitative constraints on the contribution from that broadening mechanism.
With the existing best fit of momentum broadening, we can estimate the magnetic rigidity limit which may have implications for other QCD phenomena related to the magnetic field such as synchrotron radiation~\cite{TuchinPhysRevC.82.034904}. 
Given that the single $e^+(e^-)$ momentum is much larger than the $e^+e^-$ pair momentum and the final momentum kick $k_{\perp}$ is much smaller than the single particle momentum ($p_1\simeq p_2$), $k_{\perp} \simeq \delta$ $p\simeq\delta P_{\perp}\simeq$ $e B L \simeq e B \tau_B$ where $L$ is the distance the electron travels in the magnetic field and $\tau_B$ is the duration of the magnetic field. 
Similar to the example in Ref.~\cite{TuchinPhysRevC.82.034904}, the magnetic field from this estimate is $e B \simeq k_{\perp}/L\simeq$ 14 MeV/2 fm $\simeq0.1m_{\pi}^2/\hbar$, about an order of magnitude smaller than the assumption in literature~\cite{TuchinPhysRevC.82.034904}. 
However, we emphasize that the current estimate was from the peripheral Au+Au collisions and the situation may be very different in more central collisions. 


\begin{figure*}
    \centering
    \begin{subfigure}{.50\textwidth}
      \centering
      \includegraphics[width=0.99\textwidth]{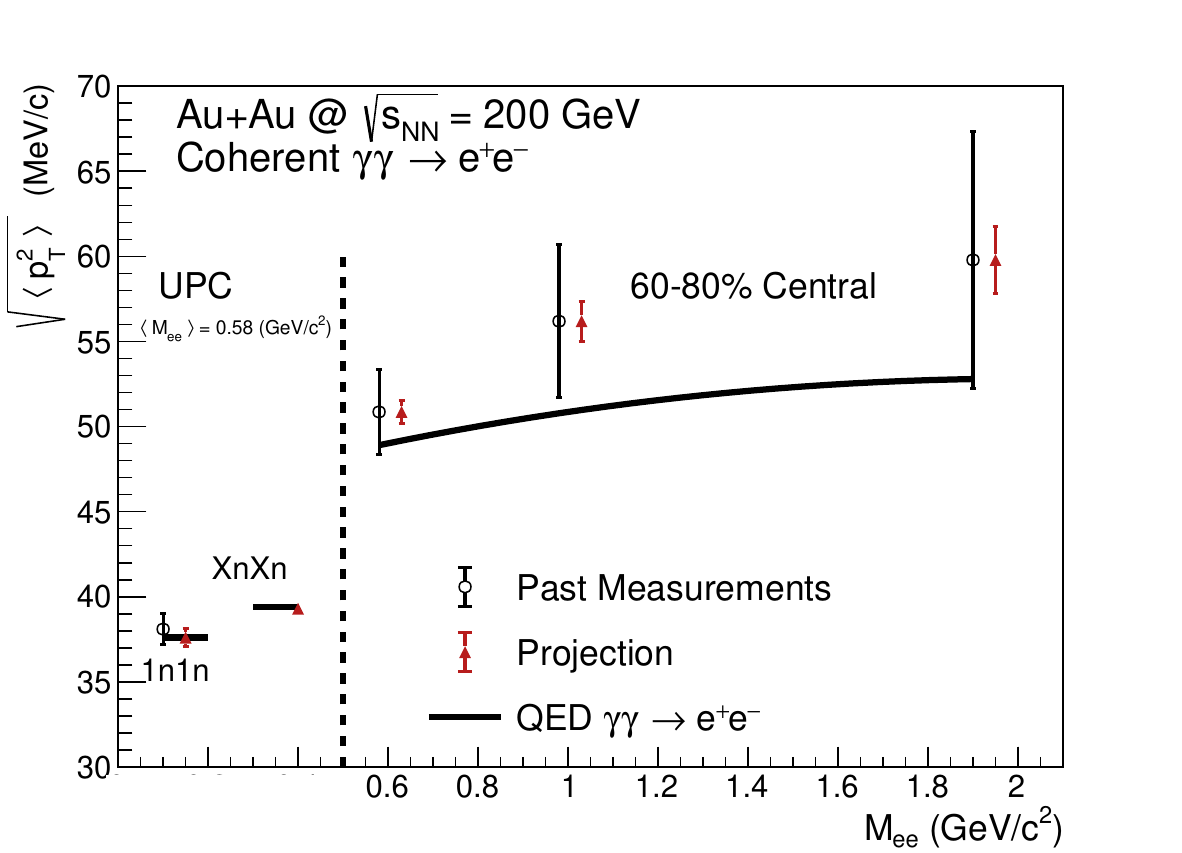}
      \caption{}
    \end{subfigure}%
    \begin{subfigure}{.50\textwidth}
      \centering
      \includegraphics[width=.99\linewidth]{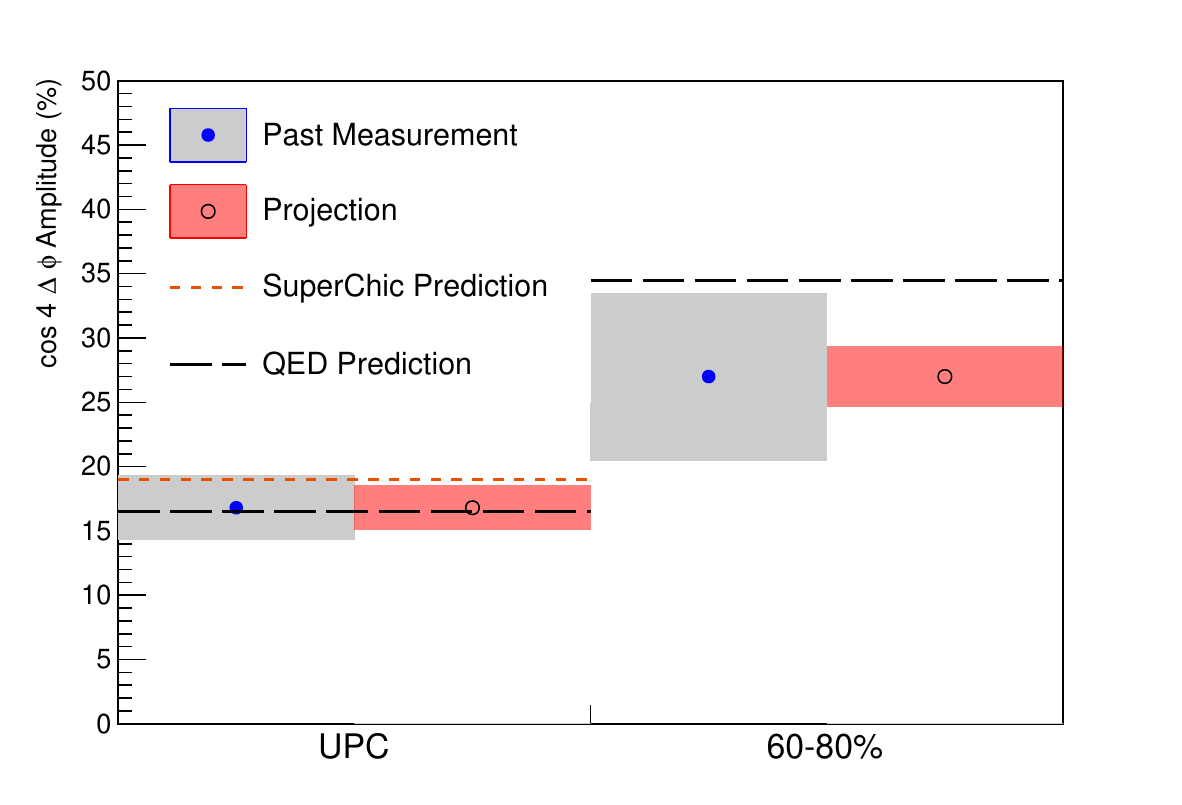}
      \caption{}
    \end{subfigure}
    \caption{(color online) \textbf{(a)} Projections for measurements of the $\gamma\gamma\rightarrow e^+e^-$ process in peripheral and ultra-peripheral collisions. Left: The $\sqrt{\langle p_T^2 \rangle }$ of dielectron pairs within the fiducial acceptance as a function of pair mass, $M_{ee}$, for $60-80\%$ central and ultra-peripheral Au+Au collisions at $\sqrt{s_{NN}}\ =$ 200 GeV. \textbf{(b)} The projection of the $\cos{4\Delta\phi}$ measurement for both peripheral $(60-80\%)$ and ultra-peripheral collisions. Reproduced from Ref.~\cite{SN0755STARBeam}. }
    \label{fig:future2}
\end{figure*}

\section{Future Opportunities}
\label{sec:future}
We close this review with a short discussion of future opportunities that are possible with additional measurements of the $\gamma\gamma$ processes in light of the collision geometry and photon polarization dependencies discussed herein. Measurement of the $\gamma\gamma$ processes in events with hadronic overlap provide the opportunity to correlate the dilepton production with other event activity. Lowest order QED calculations predict that the produced lepton pairs from the $\gamma\gamma$ processes should be preferentially correlated with the global reaction plan~\cite{zhaInitialTransversemomentumBroadening2020b} as shown in Fig.~\ref{fig:future}. Furthermore, the correlation strength is expected to change with impact parameter (centrality) range due to the changing electromagnetic field overlap distribution. 
Since the photon polarization effects lead to a $\cos4\phi$ modulation in the angle between the pair momentum $(p_{1\perp} + p_{2\perp})$ and $(p_{1\perp} - p_{2\perp})/2$, the correlation of the pair with the global event reaction plane implies that $(p_{1\perp} - p_{2\perp})/2$ is also correlated with the reaction plane. 
In this case one can identify the anisotropic effects with respect to the global magnetic field direction (approximately perpendicular to the reaction plane)~\cite{skokovEstimateMagneticField2009}.
Observation of these correlations with respect to the global event reaction plane would therefore provide novel input for the in-situ study of the ultra-strong electromagnetic fields that may lead to exotic QCD phenomena, such as the Chiral Magnetic Effect. 

As discussed in Sec.~\ref{sec:mapping}, the observation of these $\gamma\gamma$ processes in events with hadronic overlap potentially allow the purely QED processes to be used as a probe of the produced nuclear medium. Both RHIC and LHC plan future data collection that will allow high precision multi-differential analysis of these $\gamma\gamma$ processes~\cite{SN0755STARBeam,ProspectsMeasurementsPhotonInduced}. Future measurements at STAR are expected to provide significantly higher precision measurements of the $e^+e^-$ transverse momentum spectra and the $\cos4\phi$ modulation. Additionally, multi-differential measurements, such as the $\cos4\phi$ modulation strength versus pair $p_T$, will be possible. The increased precision on the pair $p_T$ will provide additional constraining power to investigate the proposed final-state broadening effects. In addition to their effect on the $p_T$ spectra, final state interactions would wash-out the $\cos4\phi$ modulation strength that results from the initial colliding photon polarization. Figure~\ref{fig:future2} shows the predicted future precision that will be achieved for the $p_T$ (a) and $\cos4\phi$ modulation measurements (b) in future STAR analyses. The added precision in the $cos4\phi$ modulation measurement is expected to allow experimental verification of impact parameter dependence predicted by the lowest order QED calculations (and therefore further exclude the $k_\perp$-factorization plus TMD treatment of the photon polarization~\cite{li_probing_2019,SuperChic3}). The future data taking campaigns planned for the LHC experiments will also allow improved measurements from ALICE of the $\gamma\gamma \rightarrow e^+e^-$ process (Fig.~\ref{fig:exp_hadronic}a) in a similar region of phase space as measured by STAR but in collisions with a much larger Lorentz-boost factor. Such measurements will provide further constraints on the treatment of the photon kinematic distributions over a range of photon energies. Similarly, future data taking by CMS and ATLAS will allow additional precision measurements of the the $\gamma\gamma \rightarrow \mu^+\mu^-$ process in events with hadronic overlap, possibly shedding light on the presence (or lack) of medium induced modifications via differential measurements of the produced dilepton kinematics.

All the proposed mechanisms discussed in this review require extraordinarily strong electromagnetic fields, an interdisciplinary subject of intense interest across many scientific communities. While much clarity has been gained in recent years, there are still a few assumptions and caveats regarding these processes which deserves further theoretical and experimental explorations~\cite{zhaInitialTransversemomentumBroadening2020b}.  In all of the theoretical approaches discussed here, there is a strong assumption of a continuous charge distribution without point-like substructure. It has been shown that the substructures of protons and quarks within nuclei~\cite{Staig:2010by} and their fluctuations~\cite{Bzdak:2011yy} can significantly alter the electromagnetic field inside the nucleus at any given instant. This should result in an observable effect, and deserves further theoretical and experimental investigation. The expected effects are most prominent in central collisions where the existing ATLAS results have large uncertainties and where STAR currently lacks the necessary statistics for a measurement. 
The very first assumption in all the known models is that both colliding nuclei maintain their velocities (a $\delta(k_{i}^{\nu}u_{i\nu})$ function) to validate the external and coherent field approximation. In central collisions, where the photon flux is expected to be generated predominantly by the participant nucleons, charge stopping and finite momentum transfer may be an important correction to the initial electromagnetic fields. 

We have ignored higher-order corrections in both the initial electromagnetic field~\cite{PhysRevC.70.031902} and the Sudakov effect~\cite{Klein:2018fmp} in this review in order to focus on the physics of the lowest order QED processes. However, the Sudakov effect may be significant at smearing the dip around $\alpha=0$, altering the high $P_{\perp}$ tail~\cite{Klein:2018fmp} and reducing the azimuthal asymmetry at high pair $P_{\perp}$~\cite{liImpactParameterDependence2020}. Furthermore, it has been pointed out that there may be, in a single event, significant multiple pair production in up to 20\% of the UPC events~\cite{Hencken:2004td} at LHC energies. This may complicate the model calculation, experimental measurements, and the corresponding comparisons in which an exact match of how the events are defined is needed between experiment and theory. Recently, various approaches for revisiting the Coulomb corrections have been proposed, in one case to the photons~\cite{Zha:2021jhf}, and in the other case to the produced $e^+e^-$ pair~\cite{Sun:2020ygb}. All these aspects need to be put into a consistent and coherent framework for comprehensive comparison to experiment and for precision tests of models based on QED~\cite{klusek-gawendaCentralityDependenceDilepton2021,Wang:2021kxm}.

In addition to the Breit-Wheeler process discussed herein, the Light-by-Light (LbyL) scattering process has also recently been observed at the LHC by the ATLAS~\cite{aaboudEvidenceLightbylightScattering2017b,atlascollaborationObservationLightbyLightScattering2019b} and CMS~\cite{sirunyanEvidenceLightbylightScattering2019a} collaborations. In the Standard Model, the leading order LbyL process proceeds through box diagrams of virtual charged particles. However, measurement of the LbyL process has also been proposed as an avenue for testing physics beyond the standard model, since the process may proceed through predicted axion-like (ALP, $a$) particles ($\gamma\gamma \rightarrow a \rightarrow \gamma\gamma$)~\cite{bauerColliderProbesAxionlike2017}. Future LHC measurements of the LbyL cross section are expected to provide additional constraining power on the allowed ALP mass and coupling~\cite{ProspectsMeasurementsPhotonInduced}. The ALP process is connected to the electromagnetic field distribution through the Lagrangian density of the form~\cite{bauerColliderProbesAxionlike2017}
\begin{align}
  \mathcal{L}_{a\gamma\gamma} = \frac{1}{4\Lambda}a F^{\mu\nu}\tilde{F}_{\mu\nu} = \frac{1}{\Lambda}a \bm{E}\cdot\bm{B},
\end{align}
where $1/\Lambda$ is the coupling strength of the interaction ($\Lambda$ has units of energy) and $a$ is the field for a massive scalar ALP. The physical link to the EM fields through the $\bm{E}\cdot\bm{B}$ term should specifically be noted. 
For this reason, a differential measurement of the LbyL cross section sensitive to the field configuration (photon polarization) could provide additional power for differentiation between the standard model process and the possible ALP mediated LbyL process. For instance, an observable similar to that used for the $\cos4\phi$ measurement performed for the Breit-Wheeler process~\cite{liImpactParameterDependence2020,liProbingLinearPolarization2019,starcollaborationMeasurementMomentumAngular2021} could be employed. In this case, $\phi = \phi_{\gamma\gamma}$ is instead the azimuthal angle between ($p_\perp^{\gamma_1} + p_\perp^{\gamma_2}$) and $(p^{\gamma_1}_\perp - p^{\gamma_2}_\perp)/2$, where $p_\perp^{\gamma_1,2}$ are the transverse momenta of the two outgoing photons. Though such a measurement may be challenging considering realistic experimental uncertainties on the measured $\gamma$ kinematics. As an example of this, we note that the SuperChic3 model displays a $\cos2\phi$ modulation in this quantity due to the photon polarization dependence~\cite{SuperChic3}. Considering the discussion above concerning the expected correlation of $l^+l^-$ pairs with the global reaction plane produced in events with hadronic overlap, one can similarly expect that the outgoing photons produced in LbyL scattering may be correlated with the global event plane as well. Therefore, if LbyL can be measured in events with hadronic overlap (thus allowing reaction plane measurement), then the angular correlation effects resulting from the photon polarization may be alternatively accessed in that way. We note that both the Breit-Wheeler process and the LbyL scattering measured in the current experiments are at midrapidity while the optical theorem connecting the two processes applies to the Breit-Wheeler process at midrapidity and the LbyL scattering at forward angles. 

As a final point about future opportunities we note that these developments have implications for the study of other photon mediated processes in heavy-ion collisions. For example, the measurement of coherent diffractive photo-nuclear production of vector mesons ($\gamma {\rm A} \rightarrow \rho^0, \phi, J/\psi ...$) has long been identified as a tool for imaging the nucleus in high energy collisions. Since these processes also involve a photon manifested from the ultra-Lorentz boosted Coulomb fields, the detailed understanding of the photon polarization and spacial dependencies that have been learned from studying the $\gamma\gamma$ processes are expected to shed new light on these other photon-mediated processes as well. Naturally, one might expect that novel phenomena may result from the impact parameter dependence and photon polarization in photo-nuclear processes~\cite{zhaExploringDoubleslitInterference2021} as have in the $\gamma\gamma$ processes. 

Since ultra-relativistic high-Z nuclei produce ultra-strong electromagnetic fields, heavy-ion collisions provide a unique experimental setting for studying QED phenomena in strong fields. In this review we have presented an overview of the concept of pair production in vacuum by strong fields followed by a brief review of the equivalent photon approximation as it is traditionally used to describe ultra-peripheral heavy-ion collisions. Recent theoretical and experimental advances have been discussed with special attention given to the recent progress in understanding the effects of impact parameter dependence and photon polarization on the kinematics of the Breit-Wheeler process. Experimental measurement of the strength and spatial distribution the ultra-strong EM fields produced in heavy-ion collisions are possible for the first time due to these advances. This progress is expected to advance the study of other areas of QED in strong fields and to provide novel input for the study of emergent phenomena of QCD in ultra-strong electromagnetic fields.

\begin{acknowledgements}
The authors would like to thank Prof. Jian Zhou, Bowen Xiao, Zebo Tang, Wei Li, Dr. Lijuan Ruan, Spencer Klein, Shuai Yang, and Chi Yang for their stimulating discussion. 
This work was funded by the National Natural Science Foundation of China
under Grant Nos. 11775213 and 11675168, 
the U.S. DOE Office of Science under contract No. de-sc0012704, DE-FG02-10ER41666, and DE-AC02-98CH10886, 
DOE Brookhaven National Laboratory LDRD 18-037, and by MOST under Grant No. 2016YFE0104800.
\end{acknowledgements}

\bibliographystyle{spphys}       

\bibliography{biblio,qedaco,special,nm}

\end{document}